\documentclass[aps,twocolumn,superscriptaddress,groupedaddress]{revtex4}  
\usepackage{graphicx}  
\usepackage{dcolumn}   
\usepackage{bm}        
\usepackage{amssymb}   
\usepackage{amsmath}
\usepackage{mathtools}
\usepackage{appendix} 
\usepackage{comment}
\usepackage{cleveref}
\usepackage[normalem]{ulem}
\usepackage{caption}
\usepackage{subcaption}

\usepackage{color}
\usepackage{comment}

\begin{document}



\title{Uncertainty Quantification for Neutrino Opacities in Core-Collapse Supernovae and Neutron Star Mergers}
\author{Zidu Lin }
\affiliation{Department of Physics and Astronomy, University of Tennessee Knoxville}

\author{Andrew W. Steiner}
\affiliation{Department of Physics and Astronomy, University of Tennessee Knoxville}
\affiliation{Physics Division, Oak Ridge National Laboratory}

\author{Jérôme Margueron}
\affiliation{Univ Lyon, Univ Claude Bernard Lyon 1, CNRS/IN2P3, IP2I Lyon, UMR 5822, F-69622, Villeurbanne, France}

\date{\today}

\begin{abstract}
We perform an extensive study of the correlations between the neutrino-nucleon inverse mean free paths (IMFPs) and the underlying equation of states (EoSs). Strong interaction
uncertainties in the neutrino mean free path are investigated in different density regimes. The nucleon effective mass, the nucleon chemical potentials, and the residual interactions in the medium play an important role in determining neutrino-nucleon interactions in a density-dependent manner. We study how the above quantities are constrained by an EoS consistent with (i) nuclear mass measurements, (ii) proton-proton scattering phase shifts, and (iii) neutron star observations. We then study the uncertainties of both the charged current and the neutral current neutrino-nucleon inverse mean free paths due to the variation of these quantities, using Hartree-Fock+random phase approximation method. Finally, we calculate the Pearson correlation coefficients between (i) the EoS-based quantities and the EoS-based quantities; (ii) the EoS-based quantities and the IMFPs; (iii) the IMFPs and the IMFPs. 
We find a strong impact of residual interactions on neutrino opacity in the spin and spin-isospin channels, which are not well constrained by current nuclear modelings.

\end{abstract}

\maketitle

\section{Introduction}
More than $98\% $ of gravitational binding energy of proto-neutron stars (PNSs) is emitted in the form of neutrinos and anti-neutrinos issued from electron captures and proton decays during the explosion of core-collapse supernovae (CCSNe). The neutrino opacity of hot and dense matter plays an important role in CCSNe explosion mechanism, particularly because the kinetic energy of the explosion is small compared with the total energy released.~\cite{Janka:2012wk,Burrows:2012ew,Mezzacappa:2005ju}. It also heavily influences the nucleosynthesis process in the neutrino driven wind (NDW)~\cite{roberts2012new,Roberts:2012um,Roberts:2016mwj}. 

The neutrino reactions in CCSNe matter can be mainly classified into two types: (1) the neutral current (NC) neutrino interactions; and (2) the charged current (CC) neutrino interactions. NC interactions are flavor-blind. Consequently, the NC neutrino opacities are similar for different flavors of neutrinos. On the other hand, the major scource of CC neutrino opacities in CCSNe matter are the $\nu_e$/$\bar{\nu}_e$ absorption/emission reactions. The CC reactions involving $\nu_{\mu}$, $\nu_{\tau}$, $\bar{\nu}_{\mu}$ and $\bar{\nu}_{\tau}$ are 
suppressed by the mass of the heavy leptons $\mu$ and $\tau$. The NC neutrino scattering reactions may have an influence on the neutrino cooling rate, the proto neutron star contraction speed in CCSNe, and the neutrino re-heating in the external layers of CCSNe~\cite{OConnor:2017ftn}. The CC neutrino absorption/emission reactions determine the $\nu_e$/$\bar{\nu}_e$ spectrum and thus the electron fraction in NDW~\cite{Roberts:2012um,roberts2012new,Roberts:2016mwj,Rrapaj:2014yba}.

Neutrino interaction rates in CCSNe matter are 
sensitive to the many-body correlations in dense and hot matter as well as in finite nuclei, which still exist in the external layers of the CCSNe close to the neutrino-sphere~\cite{Horowitz:2012us}. Close to the neutrino-sphere, the neutrino mean free path is comparable to the size of protoneutron star and the neutrino transport properties outside of this region can be well described by free streaming. Pioneering works on neutrino opacities in dense and hot matter use Hartree-Fock (HF) or HF+random phase approximations (HF+RPA) to estimate the many-body corrections on neutrino-nucleon interactions, in both the non-relativistic limit, see  Refs.~\cite{Navarro1999,Reddy:1997yr,Reddy:1998hb,Burrows:1998cg,Sawyer:1975js,Sawyer:1989nu}, and the relativistic limit, see Refs.~\cite{Reddy:1997yr,Reddy:1998hb,Roberts:2016mwj}.

In the long-wavelength limit~\cite{Horowitz:2016gul}, the many-body corrections on NC neutrino-nucleon interactions are solely determined by its equation of state (EoS). Recent progress on the description of NC neutrino-nucleon scattering rates include works using a virial equation of state (EoS) to calculate the inverse mean free path (IMPF) near CCSNe neutrino sphere model-independently~\cite{Horowitz:2016gul,Horowitz:2006pj}. As density increases, the virial expansion method gradually loses its predictive power because of our lack of understanding of the higher order virial coefficients. Early efforts in calculating NC neutrino-nucleon interactions in the high-density regime include works using HF or HF+RPA approximations~\cite{Reddy:1997yr,Reddy:1998hb,Horowitz:2006pj,Horowitz:2016gul,Sawyer:1975js,Sawyer:1989nu}. Additionally, recent works based on lattice effective field (EFT) theory provide an \emph{ab-initio} calculation of the static structure factor of neutron matter over a wide range of densities at finite temperatures~\cite{Alexandru:2019gmp,Alexandru:2020zti}. The results from lattice EFT calculations agree with the those based on the virial method in the low density regime and may give insight into the many-body corrections to NC neutrino interactions in high density regime.

In CC neutrino-nucleon reactions, the transferred neutrino energy and momentum can be much larger than those in the NC reactions, since they are governed by the in-medium single-particle energy difference between the neutrons and the protons. This difference is determined by the symmetry energy, which is expected to be larger than 30~MeV in dense matter above saturation density~\cite{Baldo:2016jhp}. Consequently, both the static and the dynamic response of the many-nucleon system are important for understanding the many-body corrections to CC neutrino interactions. To our knowledge, no model-independent description of CC reactions have been 
performed in the context of CCSNe. Early efforts calculating CC reactions include~\cite{Reddy:1997yr,Reddy:1998hb,Horowitz:2012us,Hernandez:1999zz}, where HF or HF+RPA approximations were applied. 

As discussed in the pioneering works, see Refs.~\cite{Reddy:1998hb,Burrows:1998cg,Navarro1999}, on neutrino-dense matter interactions based on HF+RPA calculations, the main source of neutrino reaction rates uncertainties is the poorly-constrained density-dependent nuclear interactions. Indeed, the density dependency of nuclear residual interactions were ignored in some early endeavors using HF+RPA to calculate neutrino-nucleon interaction rates, and the uncertainties of neutrino-nucleon interactions due to the uncertainties of nuclear interactions have not been systematically studied.  

In this work, we apply HF+RPA to calculate the dynamic and the static responses of nucleons to neutrinos in CC and NC reactions. We improve the description of the density-dependent residual interactions by 
deriving them from the virial model at low densities and from the Skyrme models at higher densities, where they are expected to be constrained by astrophysical observations, nuclear experiments, and guided from fundamental theory. Note that we limit the densities explored in this study, and consider the high-density limit to be 0.2~fm$^{-3}$. In our approach, the uncertainties of the EoS-based quantities at high densities are captured by the Skyrme models and propagate into the calculation of neutrino-nucleon reaction rates. In this work, we quantitatively study the uncertainties of neutrino opacities due to EoSs uncertainities at different densities, in the framework of HF+RPA approach. As pointed out by several CCSNe numerical simulations, neutrino opacities in different density regimes are sensitive to different CCSNe physics. By studying the \emph{density-dependent} correlations between neutrino opacities and the underlying EoSs, the critical EoS-based quantities determining the neutrino opacities may be unveiled at different densities. 

In section \ref{sec:formula}, we introduced the formalism for both CC and NC neutrino-nucleon reaction rates. In section \ref{sec:result}, we present the IMFPs, the dynamic responses of both CC and NC reactions. We also present the Pearson correlations between (1) two different EoS-based quantities; (2) EoS-based quantities and IMFPs and (3) two IMFPs taken at different densities. Finally, we conclude this analysis in section \ref{sec:conclusion}.

\section{Formalism at finite temperature}
\label{sec:formula}

The differential cross section of neutrino-nucleon reactions $l_1+N_2\rightarrow l_3+N_4$ is given by
\begin{equation}\label{eq:difcrx}
 \begin{split}
     \frac{1}{V}\frac{d^2\sigma}{dE_3d\mu_{13}}&=\dfrac{G_F^2} {4\pi^2} P_3 E_3 F_{\mathrm{Pauli}}\\&\times[\mathcal{V}^2(1+\mu_{13})S_V+\mathcal{A}^2(3-\mu_{13})S_A)],
 \end{split}
\end{equation}
where $l_{1/3}$ are the incoming/outgoing leptons, and $N_{2/4}$  are the initial/final state nucleons. The vector and axial couplings $\mathcal{V}$ and $\mathcal{A}$, in the NC reactions, stand for $C_V /2=-0.50$ and $C_A/2=-0.615$ respectively. In the CC reactions, they stand for $g_V=1$ and $g_A=1.23$ respectively. The Pauli blocking factor is $F_{\mathrm{Pauli}}=[1-f(E_3)]$ in CC reactions, where $f(E_3)$ is the Fermi distribution of final state leptons. In NC reactions, we take $F_{\mathrm{Pauli}}=1$. In mean field approximations, the response function $S_V$ and $S_A$ associated with the Fermi and Gamow-Teller operators are indistinguishable and we have $S_A=S_V=S_0$. We first discuss the response functions $S_0$ for neutrino-nucleon neutral current (NC) and charged current (CC) reactions.

\subsection{Mean Field Approximation for the Response functions}

We start from the Hartree-fock residual propagator $G_0^{\tau \tau'}$ at finite temperature defined as~\cite{FetterWalecka}:
\begin{equation}
    G_0^{\tau\tau'}(\Vec{k},q_0,\Vec{q})=\dfrac{f^{\tau}(\Vec{k})-f^{\tau'}(\Vec{k}+\Vec{q})}{q_0+\epsilon^{\tau}(\Vec{k})-\epsilon^{\tau'}(\Vec{k}+\Vec{q})+i\eta},
\end{equation}
where the neutrons and protons Fermi-Dirac distributions are
\begin{equation}
f^{\tau}(\Vec{k})=\left\{1+\exp[(\epsilon^{\tau}(\Vec{k})-\mu^{\tau})/T]\right\}^{-1} \, ,
\end{equation}
with $\tau$ and $\tau^\prime$ both refer to either neutrons or protons. The quantities $\epsilon^\tau$ are the mean-field single particle energies, see Eq.~\eqref{eq:meanfield}, and $\mu^\tau$ are the chemical potentials.
Given the propagator, the imaginary part of the polarization function $\Pi_0$, in the mean field approximation~\cite{FetterWalecka,Reddy:1997yr}, is
\begin{equation}
 \begin{split}
\mathrm{Im}~\Pi_0&=\dfrac{2 }{(2\pi)^3} \int\!\! d^3 k \, G_0(\Vec{k})\\&=\dfrac{\left\{1-\exp[(-q_0-\mu^{\tau}+\mu^{\tau'})/T]\right\}}{4\pi^2} \times\\
&\int\!\! d^3 k \, \delta(\epsilon^{\tau}-\epsilon^{\tau'}-q_0)
f^{\tau}(\Vec{k})[1-f^{\tau'}(\Vec{k}+\vec{q})] \, ,
 \end{split}\label{eq:PiMF}
\end{equation}
 The detailed balance theorem and $1/(w+ i\eta)=\mathcal{P}(1/w)-i\pi\delta(w)$ is used to derive the second line of Eq.~\eqref{eq:PiMF}. In the linear response theory, the dynamic structure factor $S_0(q_0,q)$ at finite temperature is defined as~\cite{FetterWalecka,Burrows:1998cg,Roberts:2012um}:
\begin{eqnarray}
S_0(q_0,q)&=&\dfrac{2~\mathrm{Im}\Pi_0}{1-\exp[(-q_0-\mu^{\tau}+\mu^{\tau'})/T]}\nonumber \\
&=&\dfrac{1}{2\pi^2}\int\!\! d^3 k \, \delta(\epsilon^{\tau}-\epsilon^{\tau'}-q_0)f^{\tau}(\Vec{k}) \times \nonumber \\
&& \qquad \qquad  [1-f^{\tau'}(\Vec{k}+\vec{q})] \, .
\end{eqnarray}
In non-relativistic limit of the mean field approximation, the nucleon energy spectrum $\epsilon^{\tau}$ is given as the sum of an effective-kinetic and mean-field terms:
\begin{equation}
\epsilon^{\tau}(\vec{k})=\dfrac{k^2}{2m_{\tau}^*}+U_{\tau},
\label{eq:meanfield}
\end{equation}
where $m^*_{\tau}$ is the Landau effective mass of nucleon with isospin $\tau$, and $U_\tau$ is the nucleon potential. Note that the energy delta function in $S(q_0, q)$ can be written in terms of the angle $\theta$ between $\vec{q}$ and $\vec{k}$:
\begin{eqnarray}\label{eq:q0delta}
\delta(q_0+\epsilon_{\tau}-\epsilon_{\tau'})&=&\dfrac{M^*_{\tau'}}{kq}\,\delta(\cos\theta-\cos\theta_0) \times \nonumber \\
&& \Theta(\epsilon_K^{\tau}-e_-) \, \Theta(e_+-\epsilon_K^{\tau})\,,
\end{eqnarray}
where
\begin{equation}
\cos\theta_0=\dfrac{M^*_{\tau'}}{kq}\left(c-\dfrac{\chi k^2}{2M^*_{\tau'}}\right),
\hspace{0.5cm}
\epsilon^{\tau}_K=\dfrac{k^2}{2m^*_{\tau}},
\end{equation}
and \begin{equation}\label{eq:eminus}
e_{\pm}=\dfrac{2q^2}{2\chi^2 M^*_{\tau}}\left[\left(1+\dfrac{\chi M^*_{\tau'}c}{q^2}\right)\pm\sqrt{1+\dfrac{2\chi M^*_{\tau'}c}{q^2}}\right],
\end{equation}
with $\chi=1-M^*_{\tau'}/M^*_{\tau}$ and $c=q_0+U_{\tau}-U_{\tau'}-q^2/(2M^*_{\tau'}).$

For charged current (CC) reactions, $\tau\neq\tau'$ and we focus in this work on $\nu_e+n\rightarrow p+e^-$, with $\tau=n$ and $\tau'=p$. By using the delta function~\eqref{eq:q0delta}, the $S(q_0, q)$ in the CC channel becomes
\begin{equation}
S^{CC}(q_0,q)=\dfrac{M^*_{\tau}M^*_{\tau'}T}{\pi q}\dfrac{\xi_--\xi_+}{1-\exp[-(q_0-\mu_{\tau}+\mu_{\tau'})/T]},
\end{equation}
where 
\begin{equation}\label{eq:xi}
\xi_\pm=\ln \left\{\dfrac{1+\exp[(e_{\pm}-\mu_\tau+U_\tau)/T]}{1+\exp[(e_{\pm}+q_0-\mu_{\tau'}+U_\tau)/T]}\right\}.
\end{equation}
For neutral current (NC) reactions $\tau=\tau'$, the NC $S(q_0,q)$ reduces to
\begin{equation}
    S^{NC}(q_0,q)=\dfrac{M^{*2}_{\tau}T}{\pi q}
    \left[\dfrac{q_0/T}{1-\exp(-q_0/T)}\left(1+
\dfrac{T\xi_-}{q_0}\right)\right].
\end{equation}
Note that in NC reactions $\chi\rightarrow0$. By performing a Taylor expansion of the second term in  eq. \ref{eq:eminus}, $e_-$ reduces to \cite{Reddy:1997yr} 
\begin{equation}
e_-=\dfrac{M^*_\tau}{2q^2}\left(q_0-\dfrac{q^2}{2M^*_{\tau}}\right)^2.
\end{equation}

\subsection{HF+RPA Response Functions at finite temperature}\label{sec:RPA method}

To go beyond the Hartree-Fock approximation by including the long-range correlations, we calculate the HF+RPA residual propagator and solve the Bethe-Salpeter integral equation \cite{FetterWalecka,GarciaRecio1992,Hernanhez:1997zbr,Pastore:2014aia,Dzhioev:2018ovi}:
\begin{equation}\label{eq:bethe}
\begin{split}
G_{\mathrm{RPA}}^{\alpha}(\vec{k},q_0,\vec{q})&=G_0(\vec{k},q_0,\vec{q})+G_0(\vec{k},q_0,\vec{q})\\&\hspace{-0.5cm}\times\sum_{\alpha'}\int\dfrac{d^3 k'}{(2\pi)^3}V_{ph}^{\alpha,\alpha'}(\vec{k},\vec{k}',q)G_{\mathrm{RPA}}^{\alpha'}(k',q_0,q) \, .
\end{split}
    \end{equation}
In this equation, $\alpha$ and $\alpha'$ are quantum numbers of residual pairs, e.g., in spin-isospin channels $\alpha=(S,T)$, $\vec{k}$ and $\vec{k}'$ are hole momenta, and $V_{ph}^{\alpha,\alpha'}(\vec{k},\vec{k}',q)$ is the residual interaction matrix element which describes
the RPA collective excitations of the system built on a mean field (Hartree-Fock) ground state. 

Neglecting the possible mixing between spin/isospin channels,
$V^{\alpha,\alpha'}_{ph}=\delta(\alpha,\alpha')V^{\alpha}_{ph}$, and we get the RPA polarization function as
\begin{equation}\label{eq:rpaPI}
\Pi_{\mathrm{RPA}}^{\alpha}=\dfrac{\Pi_0}{1-V_{ph}^{\alpha} \Pi_0}.
\end{equation}

In the next step, we discuss the residual interactions relevant more specifically to NC and CC neutrino-nucleon reactions.

In the monopolar Laudau approximation, where only the low-energy $\ell=0$ interaction is considered and the momenta are taken at the Fermi surface,  $V_{ph}^{\alpha}(\vec{k},\vec{k}',q)=W_1^{\alpha}+W_{1R}^{\alpha}+W^{\alpha}_2(\vec{k}_F^2+\vec{k}_F^{\prime 2})$, where $W_{1(2)}^{\alpha}$ and $W_{1R}^{\alpha}$ are the strength functions of the residual interactions. Note that $W_{1R}^{\alpha,\alpha'}$ is the contribution from the rearrangement term, i.e., the term which derives from the density dependence of the effective nuclear interaction. For Skyrme force, since the density dependent term implies only the isoscalar density $\rho$, the rearrangement term does not contribute to the spin-density channel. The $W_1^{\alpha}$, $W_{1R}^{\alpha}$  and $W_2^{\alpha}$ are strength functions, see Appendix \ref{appendix2}, which can be written in terms of the Skyrme parameters as well as of the isoscalar density $\rho$ \cite{Hernanhez:1997zbr}. The density and spin-density dependent residual interactions in $(pp^{-1},pp^{-1})$, $(nn^{-1},nn^{-1})$, $(pp^{-1},nn^{-1})$ and $(nn^{-1},pp^{-1})$ transitions are then given by
\begin{equation}\label{eq:fnn}
f_0^{\tau\tau}=\dfrac{1}{2}(W_{1}^{\tau \tau,0}+W_{1R}^{\tau \tau,0})+W_2^{\tau\tau,0}k_F(\tau)^2,
\end{equation}
\begin{equation}\label{eq:fnp}
f_0^{\tau-\tau}=\dfrac{1}{2}(W_{1}^{\tau -\tau,0}+W_{1R}^{\tau -\tau,0})+\dfrac{1}{2}W_2^{\tau -\tau,0}\left[k_F^2(\tau)+k_F^2(-\tau)\right],
\end{equation}
\begin{equation}\label{eq:gnn}
g_0^{\tau\tau}=\dfrac{1}{2}W_{1}^{\tau \tau,1}+W_2^{\tau\tau,1}k_F(\tau)^2,
\end{equation}
\begin{equation}\label{eq:gnp}
g_0^{\tau-\tau}=\dfrac{1}{2}W_{1}^{\tau -\tau,1}+\dfrac{1}{2}W_2^{\tau -\tau,1}\left[k_F^2(\tau)+k_F^2(-\tau)\right]\,.
\end{equation}
 
The strength functions $W_i^\alpha$ ($i=1$, $1R$, $2$) in symmetric nuclear matter (SNM) can be obtained from the general expression in $(\tau,\tau^\prime)$ channel $W_i^{\tau\tau',S}$ as~\cite{Hernandez:1999zz},

\begin{equation}
W_i^{0,S}=W_i^{\tau\tau,S}+W_i^{\tau-\tau,S},
\end{equation}
\begin{equation}
W_i^{1,S}=W_i^{\tau\tau,S}-W_i^{\tau-\tau,S}.
\end{equation}
from which the Landau parameters in SNM are obtained:
\begin{equation}
\label{eq:f0}
f_0=\dfrac 1 2 \left(f^{\tau \tau}_0+f_0^{\tau-\tau}\right), \hspace{0.5cm}
f_0^\prime=\dfrac 1 2 \left(f^{\tau \tau}_0-f_0^{\tau-\tau}\right)
\end{equation}
\begin{equation}
\label{eq:g0}
g_0=\dfrac 1 2 \left(g^{\tau \tau}_0+g_0^{\tau-\tau}\right), \hspace{0.5cm}
g_0^\prime=\dfrac 1 2 \left(g^{\tau \tau}_0-g_0^{\tau-\tau}\right)
\end{equation}

We now focus on the NC processes, e.g., $\nu+n\rightarrow \nu+n$, where the residual interaction $V_{ph}$ can be expressed in terms of the Landau parameters~\eqref{eq:fnn}-\eqref{eq:gnp}. The polarization functions for Fermi (Vector) and Gamow-Teller (Axial) operators are
\begin{equation}
\label{eq:NCrpaPIvec}
\Pi^{NC}_V=\dfrac{\Pi^{NC}_0}{1-f^{nn}_0 \Pi^{NC}_0} \, ,
\end{equation}
and \begin{equation}\label{eq:NCrpaPIax}
\Pi^{NC}_{A}=\dfrac{\Pi^{NC}_0}{1-g^{nn}_0 \Pi^{NC}_0} \, .
\end{equation}

In $(n.p.e)$ beta equilibrium condition, residual interactions corresponding to $(nn^{-1},pp^{-1})$ are also involved for the NC RPA calculation. The vector and the axial vector residual interactions in matrix form are written as:
\begin{equation}
V_{ph}^{V}=
\begin{bmatrix}
f_{nn}&f_{np}\\f_{pn}&f_{pp}
\end{bmatrix}\, ,
\hspace{0.5cm}
V_{ph}^{A}=
\begin{bmatrix}
g_{nn}&g_{np}\\g_{pn}&g_{pp}
\end{bmatrix}\, .
\end{equation}

Furthermore, the Hartree-Fork polarization function may be written as a $2\times 2$ diagonal matrix. We then get \cite{Reddy:1998hb}
\begin{equation}
\begin{split}
    \begin{bmatrix}
    \Pi_{\mathrm{V(A)}}^{nn}&\Pi_{\mathrm{V(A)}}^{np}\\ \Pi_{\mathrm{V(A)}}^{pn}&\Pi_{\mathrm{V(A)}}^{pp}
    \end{bmatrix}&=\begin{bmatrix}
    \Pi_0^{n}&0\\ 0&\Pi_0^{p}
    \end{bmatrix}\\&+\begin{bmatrix}\Pi_{\mathrm{V(A)}}^{nn}&\Pi_{\mathrm{V(A)}}^{np}\\ \Pi_{\mathrm{V(A)}}^{pn}&\Pi_{\mathrm{V(A)}}^{pp}\end{bmatrix}V_{ph}^{V(A)}\begin{bmatrix}
    \Pi_0^{n}&0\\ 0&\Pi_0^{p}
    \end{bmatrix}.
\end{split}
\end{equation}
A compact form of RPA polarization functions including coupling constants is
\begin{eqnarray}\label{eq:polfunction}
C^T\Pi_{\mathrm{V(A)}}^{NC}C &=&
c^2_{n,V(A)}\Pi^{nn,V(A)}_{\mathrm{V(A)}}+c^2_{p,V(A)}
\Pi^{pp,V(A)}_{\mathrm{V(A)}}\nonumber \\
&& +2c_{n,V(A)}c_{p,V(A)}
\Pi^{np,V(A)}_{\mathrm{V(A)}},
\end{eqnarray}
where 
\begin{equation}
    C=\begin{bmatrix}
c_{n,V(A)}\\c_{p,V(A)}
\end{bmatrix}
\end{equation} are the coupling constants. When calculating NC vector current neutrino-proton interactions, we drop the second and the third term in the right handed side of Eq.~\eqref{eq:polfunction}, which are proportional to the vector current neutrino-proton coupling constant $c_{p,V}\approx0$. The RPA polarization function in NC vector current channel, is
\begin{equation}\label{eq:ncvecPI}
\begin{split}
    \Pi_{\mathrm{V}}^{NC}=\dfrac{(1-f_{pp}\Pi_{0}^p)\Pi_{0}^{n}}{\Delta_V} \, .
\end{split}
\end{equation}
For
the neutral current vector polarization functions, we have 
\begin{equation}
    \Delta_V=1-f_{nn}\Pi_{0}^{n}-f_{pp}\Pi_{0}^p+f_{pp}\Pi_{0}^pf_{nn}\Pi_{0}^n-f_{np}^2\Pi_{0}^p\Pi_{0}^n \, .
\end{equation} 
Note that the effect of Coulomb force in the NC vector polarization functions has been considered, by replacing $f_{pp}$ with $f_{pp}+4\pi e^2(q^2+q_{TF}^2)^{-1}$, where $q_{TF}=4e^2\pi^{1/3}(3n_p)^{2/3}$. 

The axial vector polarization function is given by \begin{equation}\label{eq:ncaxPI}
    \Pi_{\mathrm{A}}^{NC}=\dfrac{(1-g_{nn}\Pi_{0}^n)\Pi_{0}^p+(1-g_{pp}\Pi_{0}^p)\Pi_{0}^n-2g_{np}\Pi_{0}^n\Pi_{0}^p}{\Delta_A},
\end{equation}
where \begin{equation}
    \Delta_A=1-g_{nn}\Pi_{0}^{n}-g_{pp}\Pi_{0}^p+g_{pp}\Pi_{0}^pg_{nn}\Pi_{0}^n-g_{np}^2\Pi_{0}^p\Pi_{0}^n.
\end{equation} 
The NC polarization functions have the form similar to those derived in Ref.~\cite{Reddy:1998hb}. Additionally, note that in SNM, we have $f_{nn}=f_{pp}=f_0+f_0'$ and $g_{nn}=g_{pp}=g_0+g_0'$. By replacing $f_{nn}$ and $f_{pp}$ with $f_0+f_0'$, Eq.~\eqref{eq:ncvecPI}  reproduces the vector RPA polarization function in \cite{Burrows:1998cg}. And by assuming $g_{nn}=g_{pp}=-g_{np}$, Eq.~\eqref{eq:ncaxPI} reproduces the axial RPA polarization function in Ref.~\cite{Burrows:1998cg}. 

For CC, the residual interactions for charge exchange (CE) process are given by~\cite{Hernandez:1999zz}:
\begin{equation}
    V_{ph}^{CE;S}=\sum_S W_i^{CE;S}P^S,
\end{equation}
and we get the residual interactions in vector and axial vector channel 
\begin{equation}\label{eq:vf}
   V_\mathrm{f}=\dfrac{1}{2}W_1^{CE;0}+W_2^{CE;0}k_F^2, 
\end{equation}
\begin{equation}\label{eq:vgt}
   V_{\mathrm{gt}}=\dfrac{1}{2}W_1^{CE;1}+W_2^{CE;1}k_F^2 
\end{equation}
corresponding to $(pn^{-1},pn^{-1})$ transitions, where $k_F$ is the Fermi momentum of holes states. The functions $W^{CE;S}$ for $S=0$ and 1 are given in terms of the Skyrme parameters: 
\begin{equation}
    W_1^{CE;0}(q=0)=-t_0(1+2x_0)-\dfrac{1}{6}t_3\rho^\gamma(1+2x_3),
\end{equation}
\begin{equation}
    W_2^{CE;0}(q=0)=-\dfrac{1}{4}[t_1(1+2x_1)-t_2(1+2x_2)],
\end{equation}
\begin{equation}
    W_1^{CE;1}(q=0)=-t_0-\dfrac{1}{6}t_3\rho^\gamma,
\end{equation}
\begin{equation}
    W_2^{CE;1}(q=0)=-\dfrac{1}{4}(t_1-t_2).
\end{equation}

In SNM, $f_0^{\mathrm{CE}}=2f_0'$ and $g_0^{\mathrm{CE}}=2g_0'$, which is consistent with the CC residual interactions used in \cite{Reddy:1998hb}. The relationship between CC and NC residual interactions is discussed in more details in Appendix \ref{appendix1}. The Landau parameters in Eqs.~\eqref{eq:vf}-\eqref{eq:vgt} are relevant to $V_{ph}$ in CC process. Indeed, in $\nu_e+n\rightarrow e^-+p$, the polarization function for Fermi and Gammow-Teller operators are:
\begin{equation}\label{eq:CCrpaPIvec}
    \Pi^{CC}_V=\dfrac{\Pi^{CC}_{0}}{1-V_\mathrm{f} \Pi^{CC}_{0}},
\end{equation}
and \begin{equation}\label{eq:CCrpaPIax}
    \Pi^{CC}_A=\dfrac{\Pi^{CC}_{0}}{1-V_\mathrm{gt} \Pi^{CC}_{0}}.
\end{equation}

In low density region where the nucleon gas is hot and dilute, we use the virial EoS \cite{Horowitz:2016gul} to deduce the residual interactions. In the Virial EoS, we have nucleon densities and pressures written in terms of spin and isospin dependent fugacities $z_i=e^{\mu_i/T}$, where $z_i$ could be $z_{n\uparrow}$, $z_{n\downarrow}$, $z_{p\uparrow}$ or $z_{p\downarrow}$. The thermal wavelength $\lambda=(2\pi/(mT))^{1/2}$, the spin-like virial coefficients are $b_1$, and the spin-opposite virial coefficients are $b_0$. The pressure in virial expansion reads:
\begin{equation}
    \begin{split}
        P&=\dfrac{T}{\lambda^3}(z_{n\uparrow}+z_{n\downarrow}+z_{p\uparrow}+z_{p\downarrow}\\&+b_{n,1}(z_{n\uparrow}^2+z_{n\downarrow}^2+z_{p\uparrow}^2+z_{p\downarrow}^2)+2b_{n,0}(z_{n\uparrow}z_{n\downarrow}+z_{p\uparrow}z_{p\downarrow})\\&+2b_{pn,1}(z_{n\uparrow}z_{p\uparrow}+z_{n\downarrow}z_{p\downarrow})+2b_{pn,0}(z_{n\uparrow}z_{p\downarrow}+z_{n\downarrow}z_{p\uparrow})).
    \end{split}
\end{equation}
Given the pressure, the nucleon density of specified spin and isospin can be derived using
\begin{equation}\label{eq:densityfugacity}
    n_{i}=\dfrac{1}{T}\dfrac{\partial P}{\partial z_{i}}z_{i}.
\end{equation}
We further write down the free energy in terms of the virial coefficients $b_i$ and the fugacities $z_i$:
\begin{equation}
\begin{split}
    f&=n_n\mu_n+n_p\mu_p-P\\&=\dfrac{T}{2}(n_{n\uparrow}\ln z_{n\uparrow}+n_{n\downarrow}\ln z_{n\downarrow}+n_{n\uparrow}\ln z_{n\downarrow}+n_{n\downarrow}\ln z_{n\uparrow}\\&+n_{p\uparrow}\ln z_{p\uparrow}+n_{p\downarrow}\ln z_{p\downarrow}+n_{p\uparrow}\ln z_{p\downarrow}+n_{p\downarrow}\ln z_{p\uparrow})\\&-P,
\end{split}
\end{equation}
where the second line was derived by using the fact that $n_{n(p)}=n_{n\uparrow(p\uparrow)}+n_{n\uparrow(p\uparrow)}$ and $z_{n(p)}=\sqrt{z_{n\uparrow(p\uparrow)}z_{n\downarrow(p\downarrow)}}$. Note that the free energy in non-interacting nucleon gas $f^0$ in virial expansion can be easily calculated by replacing the virial coefficients: $b_{n,1}\rightarrow b_{n,1}^{\mathrm{free}}=-1/4\sqrt{2}$, $b_{n,0}\rightarrow b_{n,0}^{\mathrm{free}}=0$, $b_{pn,1}\rightarrow b_{pn,1}^{\mathrm{free}}=0$ and $b_{pn,0}\rightarrow b_{pn,0}^{\mathrm{free}}=0$. Finally, we have the nucleon potential energy $U_i$ in virial expansion:
\begin{equation}\label{eq:virialU}
    U_i=\dfrac{\partial(f-f^0)}{\partial n_i},
\end{equation}
which is defined similarly as in Ref.~\cite{Horowitz:2012us}.
Given the potential density $E=f-f_0$ and the single nucleon potential $U_i$, in principle we can derive the spin and isospin dependent residual interactions by finding the double density functional derivative of potential energy density $E$:\begin{equation}
    v_{ph}(i,j)=\dfrac{\delta^2 E}{\delta n_i \delta n_j}=\dfrac{\delta U_i}{\delta n_j},
\end{equation}
where the index $i$ indicate the spin and isospin of the nucleon states. In SNM, the monopolar Laudau parameters are defined as~\cite{Sawyer:1989nu}: \begin{equation}
  f_0=\dfrac{\partial^2 E}{\partial n\partial n},  
\end{equation}
\begin{equation}
  f'_0=\dfrac{\partial^2 E}{\partial n_{3,0}\partial n_{3,0}},  
\end{equation}
\begin{equation}
  g_0=\dfrac{\partial^2 E}{\partial n_{0,3}\partial n_{0,3}},  
\end{equation}
and
\begin{equation}
  g'_0=\dfrac{\partial^2 E}{\partial n_{3,3}\partial n_{3,3}},  
\end{equation}
where $n_{3,0}=n_p-n_n$, $n_{0,3}=n_{p\uparrow}-n_{p\downarrow}+n_{n\uparrow}-n_{n\downarrow}$ and $n_{3,3}=n_{p\uparrow}-n_{p\downarrow}-n_{n\uparrow}+n_{n\downarrow}$. Following Ref.~\cite{Horowitz:2012us}, we invert Eq.~\eqref{eq:densityfugacity} to second order in densities. In this way, the free energy is a function of densities up to the second order:
\begin{equation}\label{eq:virialf}
\begin{split}
    f&=-T(n_{n\uparrow}+n_{n\downarrow}+n_{p\uparrow}+n_{p\downarrow})-2b_{pn,0}T\lambda^3(n_{p\uparrow} n_{n\downarrow}\\&+n_{p\downarrow}n_{n\uparrow})-2b_{pn,1}T\lambda^3(n_{p\uparrow}n_{n\uparrow}+n_{p\downarrow}n_{n\downarrow})-2b_{n,0}T\lambda^3\\&(n_{n\uparrow}n_{n\downarrow}+n_{p\uparrow}n_{p\downarrow})-b_{n,1}T\lambda^3(n_{n\uparrow}^2+n_{p\uparrow}^2+n_{n\downarrow}^2+n_{p\downarrow}^2)+\\&\dfrac{T}{2}(n_{n\uparrow}\ln[\lambda^3n_{n\uparrow}]+n_{n\downarrow}\ln[\lambda^3n_{n\downarrow}]+n_{n\uparrow}\ln [\lambda^3n_{n\downarrow}]+\\&n_{n\downarrow}\ln[\lambda^3n_{n\uparrow}]+n_{p\uparrow}\ln [\lambda^3n_{p\uparrow}]+n_{p\downarrow}\ln [\lambda^3n_{p\downarrow}]\\&+n_{p\uparrow}\ln [\lambda^3n_{p\downarrow}]+n_{p\downarrow}\ln [\lambda^3n_{p\uparrow}]).
\end{split}
    \end{equation} 
Note that the nucleon potential $U_i$ in Eq.~\eqref{eq:virialU} reproduces the nucleon potentials in spin-symmetric matter in \cite{Horowitz:2012us}. Since the virial EoS include virial coefficients up to the 2nd order, $f$ is a function of $n_i$ up to $\mathcal{O}(n^2)$. Consequently, the residual interactions based on low-density virial expansion are density and isospin-density independent. They are given by:
\begin{equation}\label{eq:virialf0}
\begin{split}
     f_{0,\mathrm{virial}}&=-\dfrac{\lambda^3T}{2}(b_{n,0}-b_{n,0}^{\mathrm{free}}+b_{n,1}\\&-b_{n,1}^{\mathrm{free}}+b_{pn,0}-b_{pn,0}^{\mathrm{free}}+b_{pn,1}-b_{pn,1}^{\mathrm{free}}),
\end{split}
\end{equation}
\begin{equation}\label{eq:virialf0p}
\begin{split}
    f_{0,\mathrm{virial}}'&=-\dfrac{\lambda^3T}{2}(b_{n,0}-b_{n,0}^{\mathrm{free}}+b_{n,1}\\&-b_{n,1}^{\mathrm{free}}-b_{pn,0}+b_{pn,0}^{\mathrm{free}}-b_{pn,1}+b_{pn,1}^{\mathrm{free}}),
\end{split}
\end{equation}
\begin{equation}\label{eq:virialg0}
\begin{split}
    g_{0,\mathrm{virial}}&=\dfrac{\lambda^3T}{2}(b_{n,0}-b_{n,0}^{\mathrm{free}}-b_{n,1}+b_{n,1}^{\mathrm{free}}+b_{pn,0}\\&-b_{pn,0}^{\mathrm{free}}-b_{pn,1}+b_{pn,1}^{\mathrm{free}}),
\end{split}
\end{equation}
 \begin{equation}\label{eq:virialg0p}
 \begin{split}
      g_{0,\mathrm{virial}}'&=\dfrac{\lambda^3T}{2}(b_{n,0}-b_{n,0}^{\mathrm{free}}-b_{n,1}+b_{n,1}^{\mathrm{free}}-b_{pn,0}\\&+b_{pn,0}^{\mathrm{free}}+b_{pn,1}-b_{pn,1}^{\mathrm{free}}).
 \end{split}
\end{equation} 

Here the $b_{0,1}$ are virial coefficients for spin-opposite and spin-like particles, $b^{\mathrm{free}}$ are virial coefficients for non-interacting nucleon gas, and the length parameter $\lambda=(2\pi/(MT))^{1/2}$. The residual interactions based on virial coefficients at $\mathrm{T}=10 ~\mathrm{MeV}$ are
\begin{eqnarray}
    f_{0,\mathrm{virial}}&=&-2.03\times 10^{-4} \;\hbox{MeV}^{-2} \, , \\
    f_{0,\mathrm{virial}}'&=&1.19 \times 10^{-4} \;\hbox{MeV}^{-2} \, , \\
    g_{0,\mathrm{virial}}&=&2.88\times10^{-6} \;\hbox{MeV}^{-2} \,, \\
    g_{0,\mathrm{virial}}'&=&7.31\times10^{-5} \;\hbox{MeV}^{-2} \, .
\end{eqnarray}
 Correspondingly, the virial coefficients at $\mathrm{T}=10~\mathrm{MeV}$ are: $b_{n,0}=0.463$, $b_{n,1}=-0.152$, $b_{pn,0}=0.725$, $b_{pn,1}=1.130$, $b_{n,0}^{\mathrm{free}}=0$,$b_{n,1}^{\mathrm{free}}=-1/(4\sqrt{2})$, $b_{pn,0}^{\mathrm{free}}=b_{pn,1}^{\mathrm{free}}=0$. Because the residual interactions are density and isospin-density independent in low density regimes where the viral approximation applies, Eq.~\eqref{eq:f0}-\eqref{eq:g0} are inverted to find $f_{nn}$,$f_{np}$, $g_{nn}$ and $g_{np}$ in NC vector and axial vector RPA polarization functions in asymmetric matter here. 
 
Furthermore, 
 \begin{equation}\label{eq:vfvirial}
   V_{\mathrm{f},\mathrm{virial}}=2f_{0,\mathrm{virial}}' 
 \end{equation} and 
 \begin{equation}\label{eq:vgtvirial}
    V_{\mathrm{gt},\mathrm{virial}}=2g_{0,\mathrm{virial}}' 
 \end{equation} are residual interactions in CC vector and axial vector RPA polarization functions in low density region where virial EoS is valid. Similarly, these residual interactions follow the form of $V_\mathrm{f}$ and $V_\mathrm{gt}$ in SNM, since the residual interactions are approximately density and isospin-density independent in low density regimes where the viral approximation applies.
 
 Finally, following Ref.~\cite{PhysRevC.99.025803}, a thermodynamically consistent approach is employed to connect the residual interactions calculated by the virial approach with the ones calculated using Skyrme models. We discuss this method in more details in Appendix \ref{appendix1}.
 
 \subsection{EoS and its Constraints}
 
 \begin{figure*}[htp]
	\centering
	\includegraphics[width=0.45\textwidth]{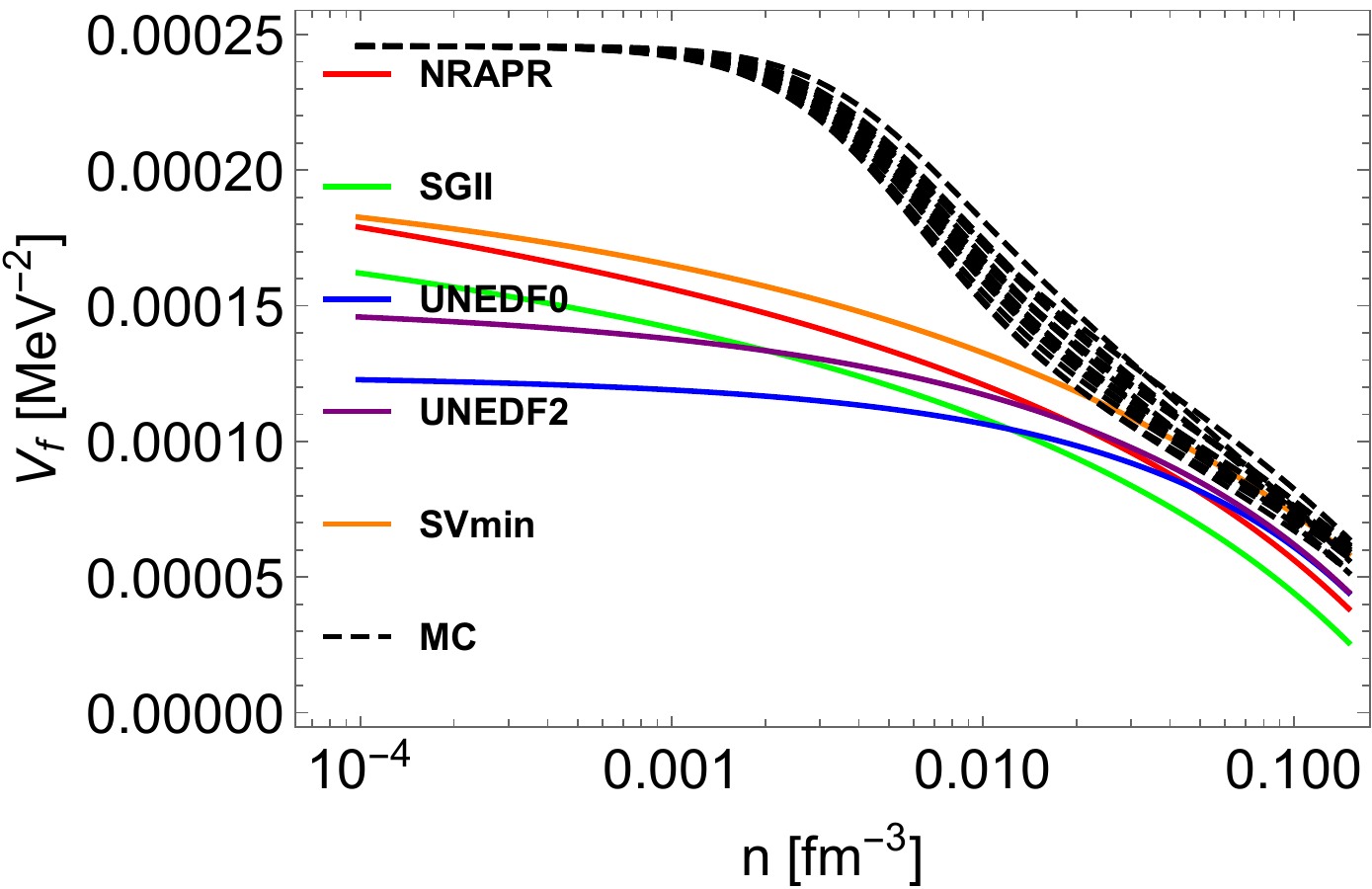}
	\includegraphics[width=0.45\textwidth]{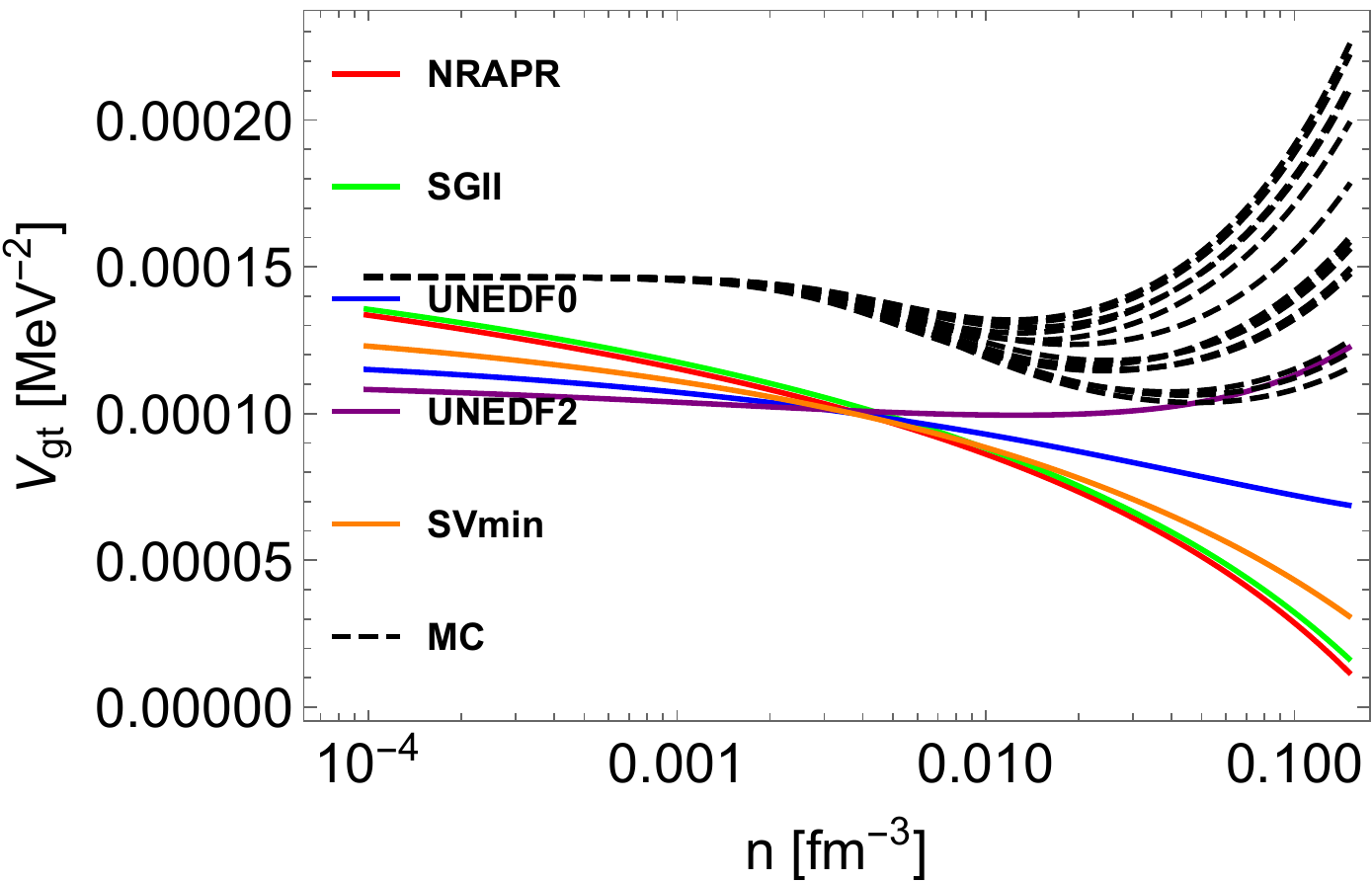}
	\includegraphics[width=0.45\textwidth]{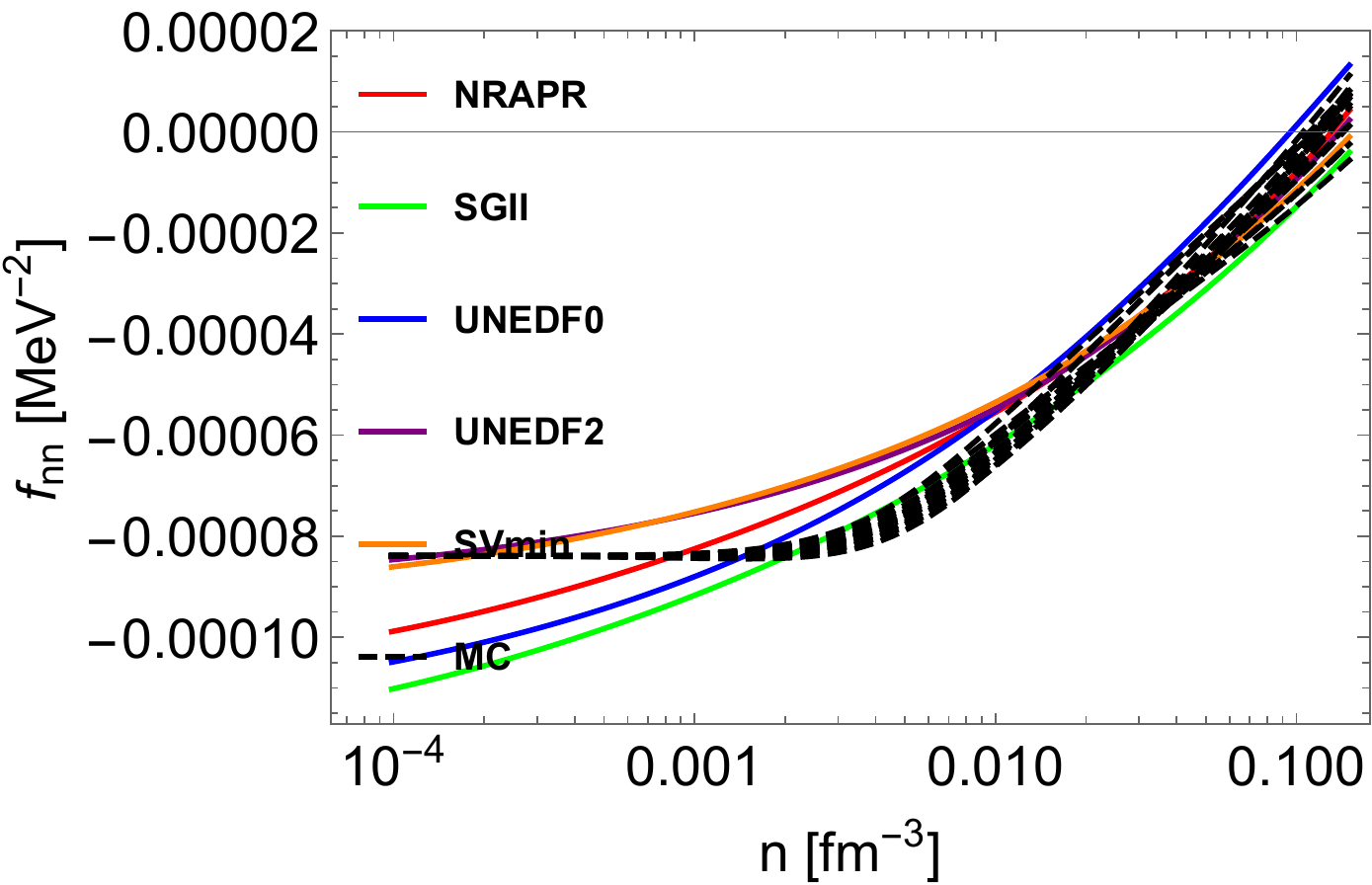}
	\includegraphics[width=0.45\textwidth]{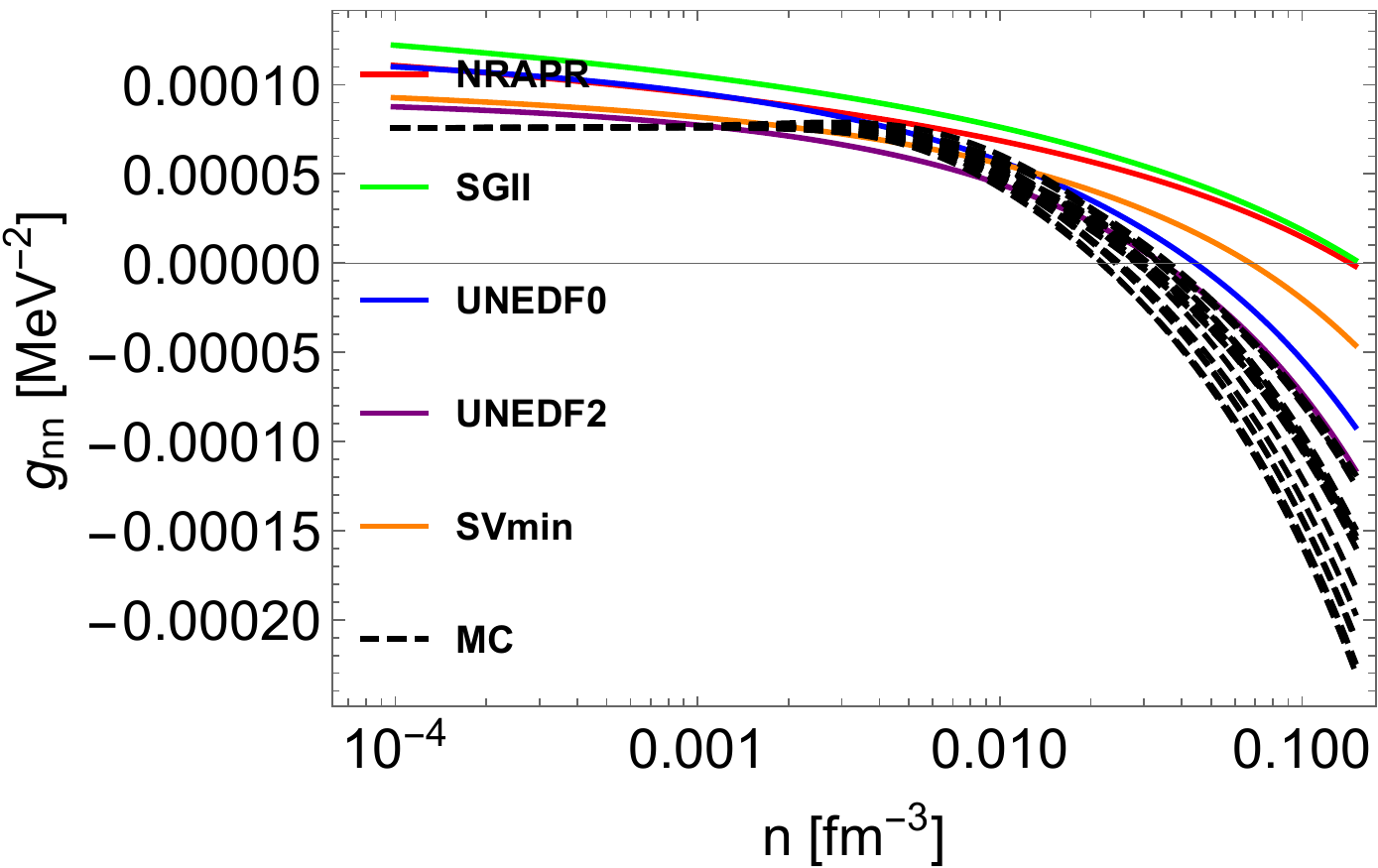}
	\includegraphics[width=0.45\textwidth]{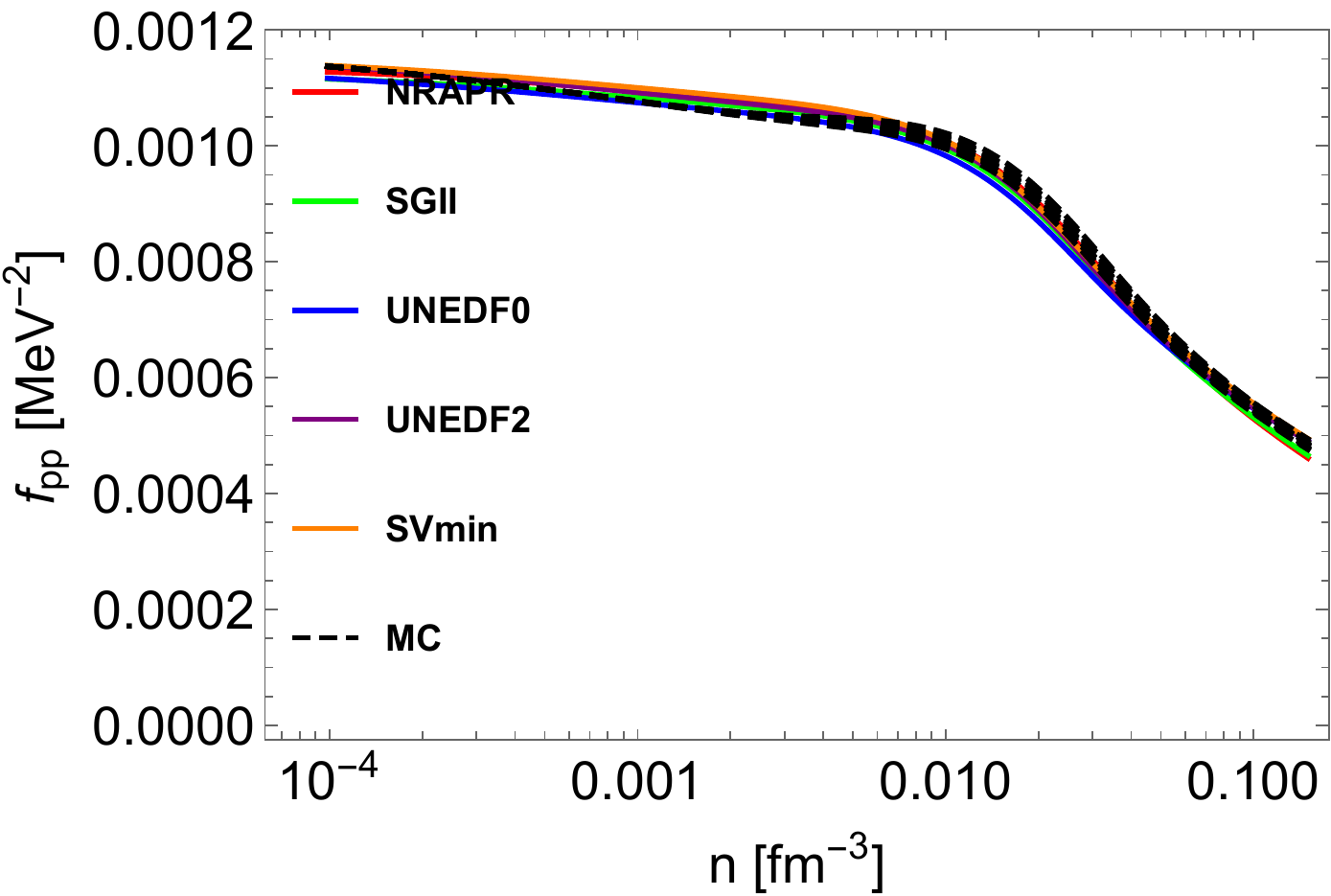}
	\includegraphics[width=0.45\textwidth]{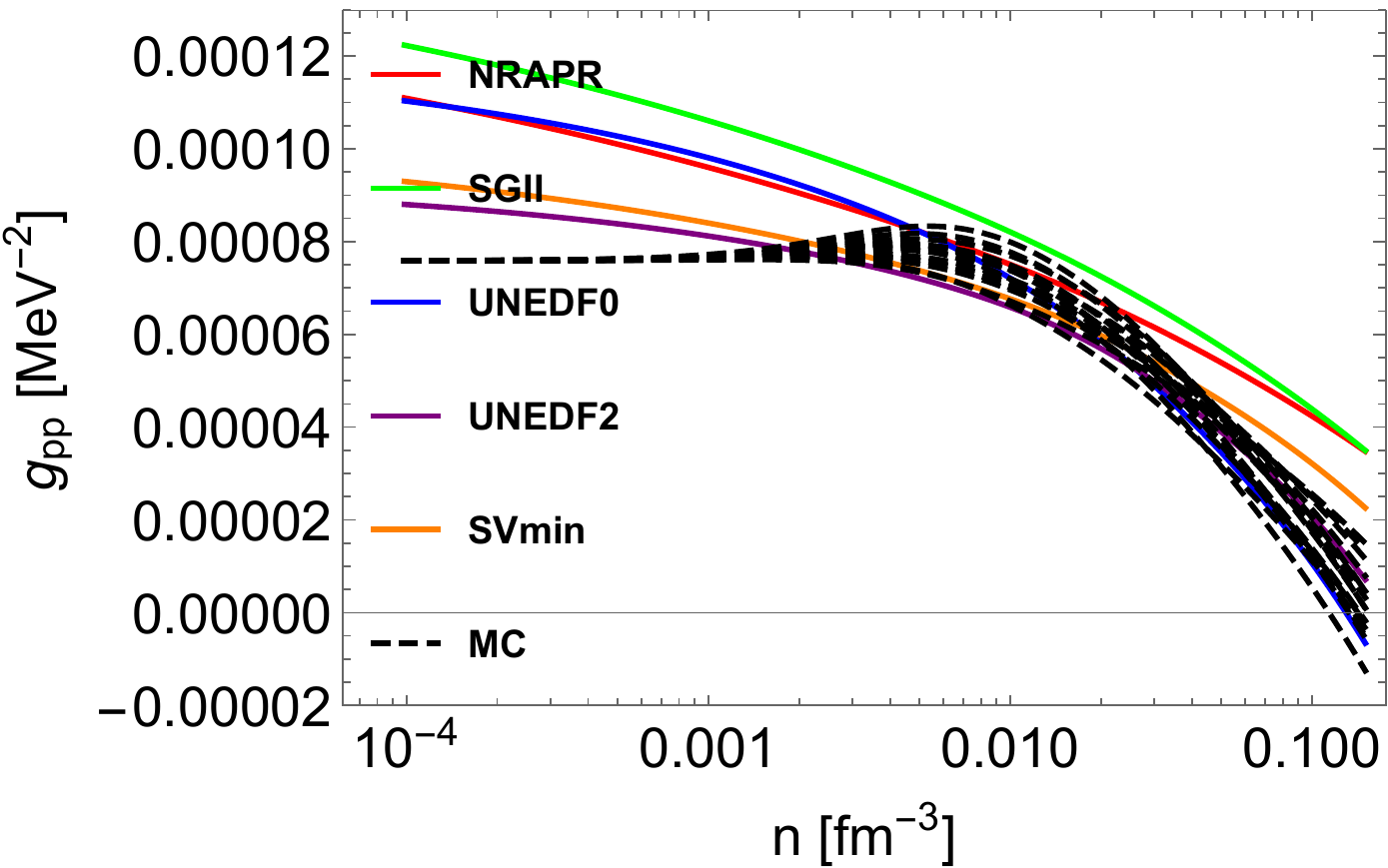}
	\includegraphics[width=0.45\textwidth]{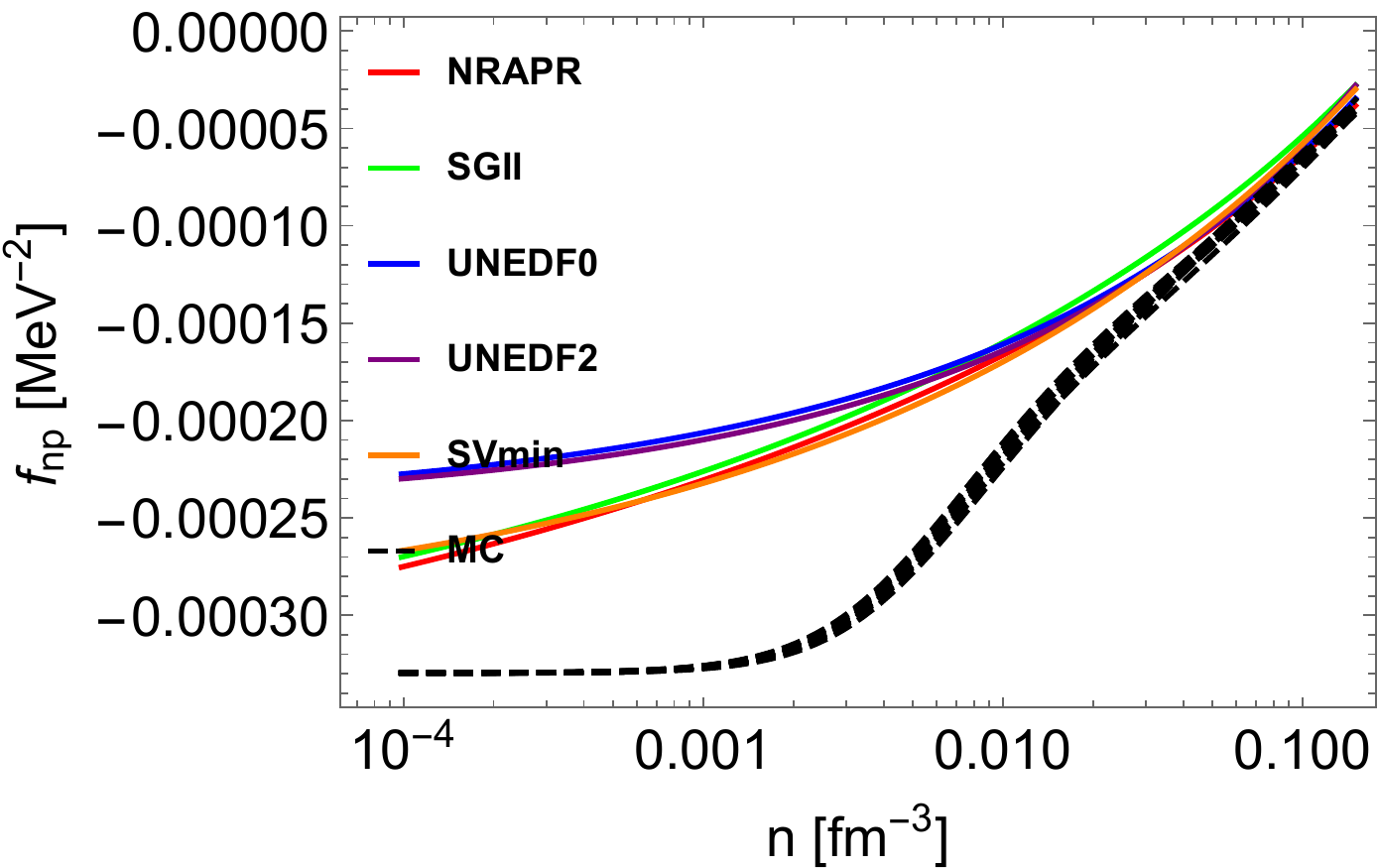}
	\includegraphics[width=0.45\textwidth]{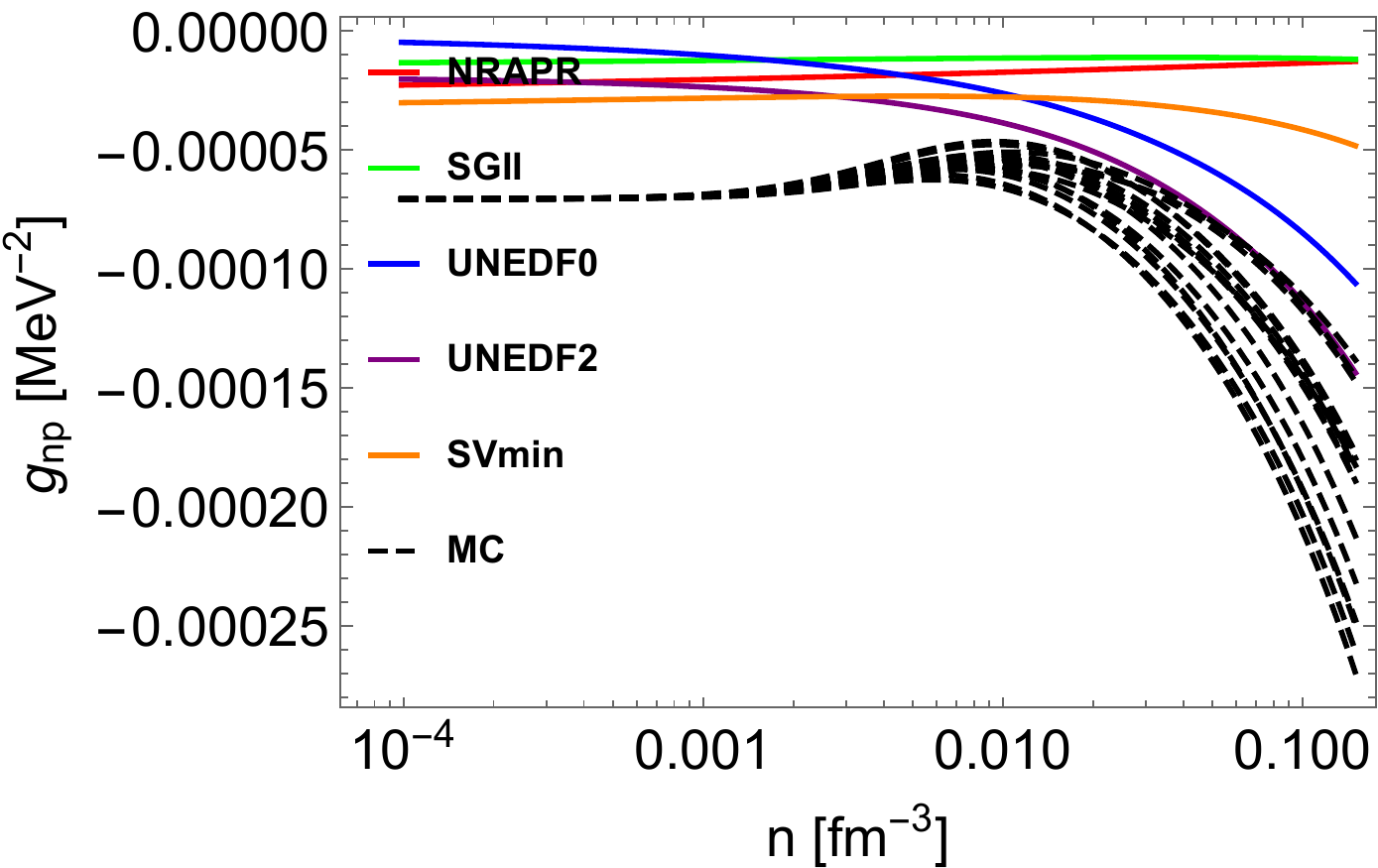}
	
	\caption{Residual interactions as function of density. In the left column, residual interactions in vector current channel are shown. In the right column, residual interactions in axial vector current channel are shown. The colored solid curves stand for residual interactions derived from Skyrme-type interactions. The black dashed curves stand for residual interactions derived from MC EoSs, see section \ref{sec:formula} for detailed description of the underlying EoSs used to derive the residual interactions. The detailed formula expression of the residual interactions in NC(CC) channel are provided in the Appendix (see Eqs. ~\eqref{eq:fnngeneral},~\eqref{eq:fnpgeneral},~\eqref{eq:gnngeneral}, and ~\eqref{eq:gnpgeneral} for NC residual interaction, and Eqs. ~\eqref{eq:vfgeneral} and ~\eqref{eq:vgtgeneral} for CC residual interactions). } 
\label{fig:phinteractions}	
\end{figure*}

\begin{figure*}[htp]
	\centering
         \includegraphics[width=0.45\textwidth]{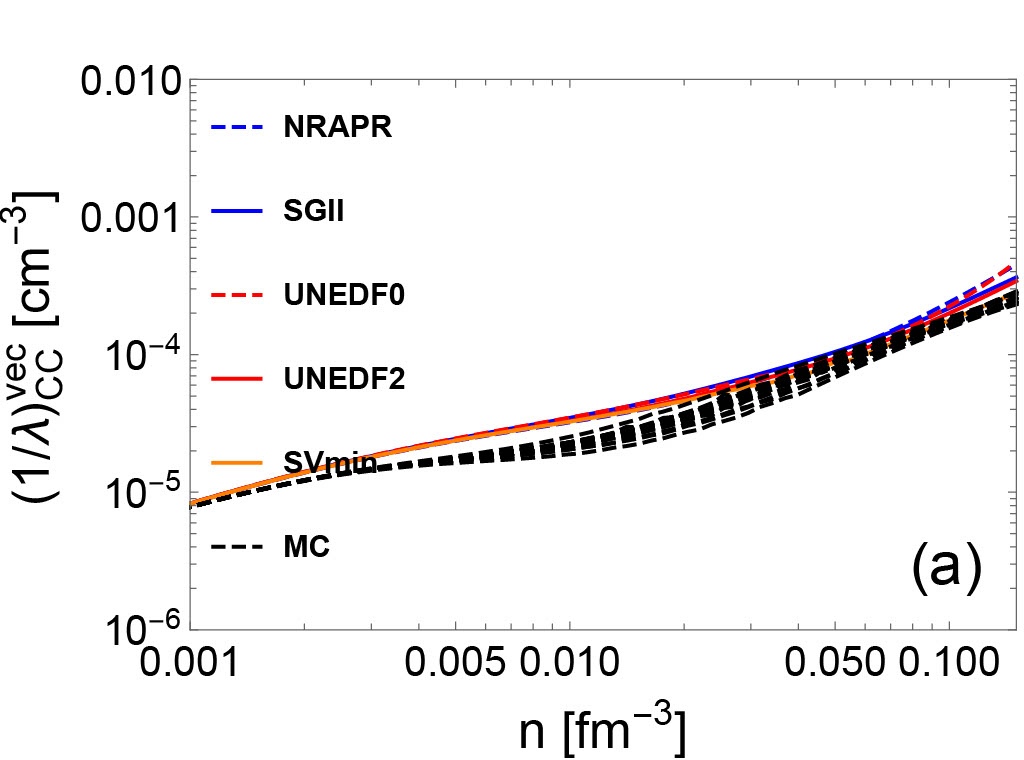} \includegraphics[width=0.45\textwidth]{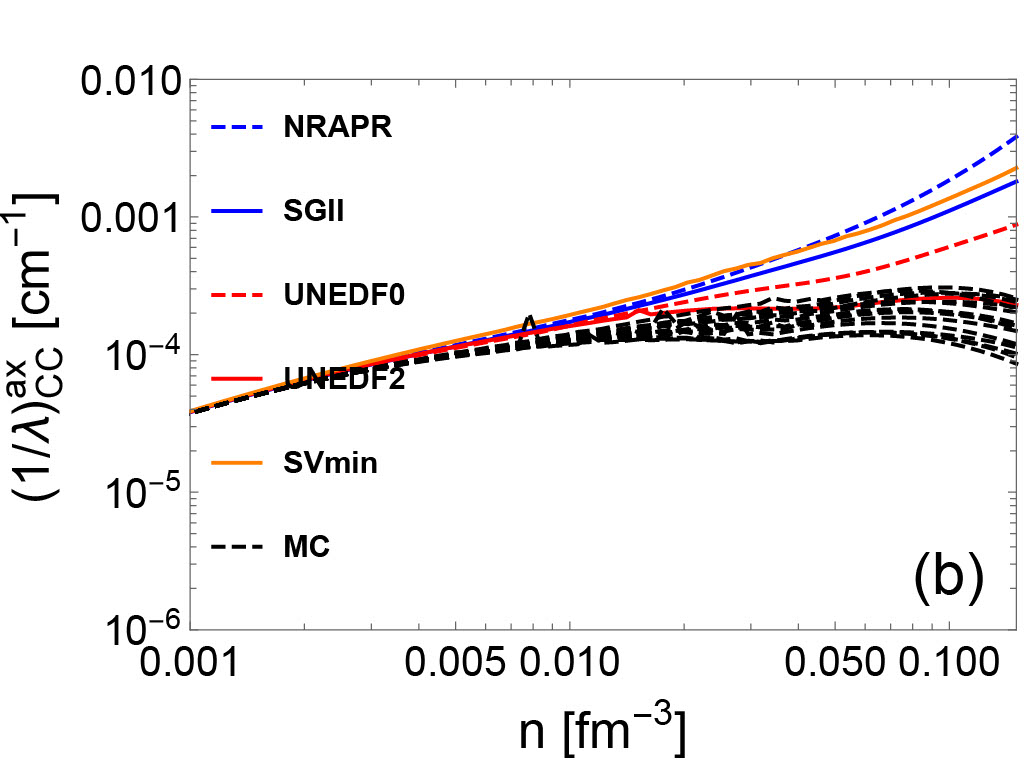}     
         \includegraphics[width=0.45\textwidth]{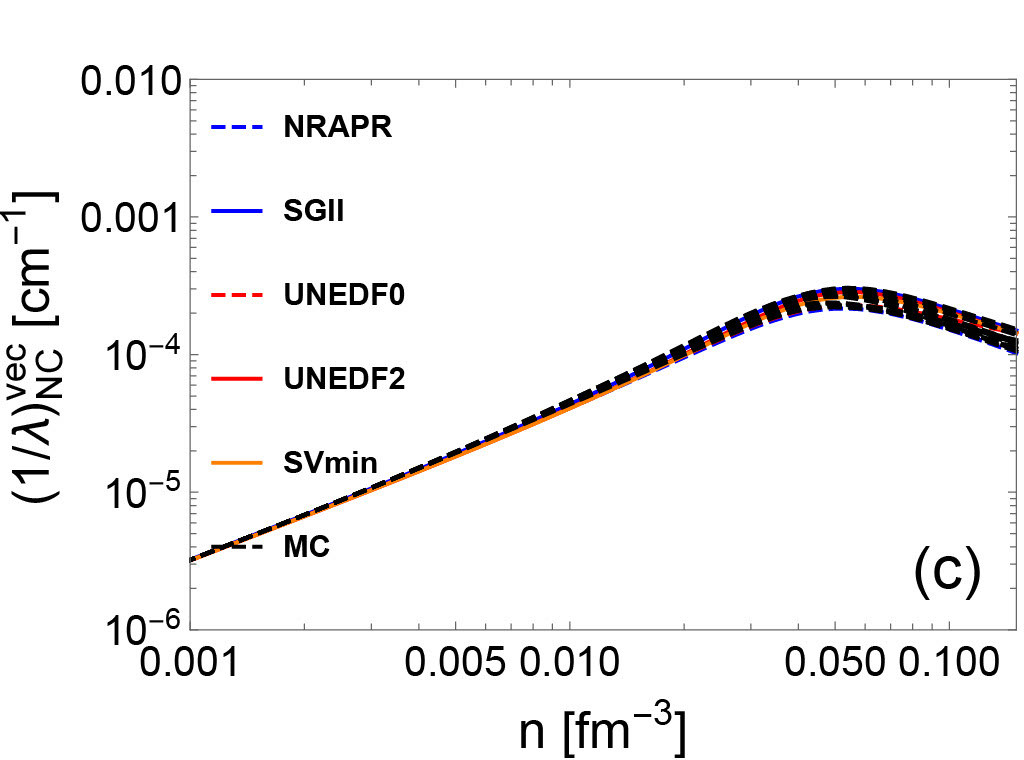}  \includegraphics[width=0.45\textwidth]{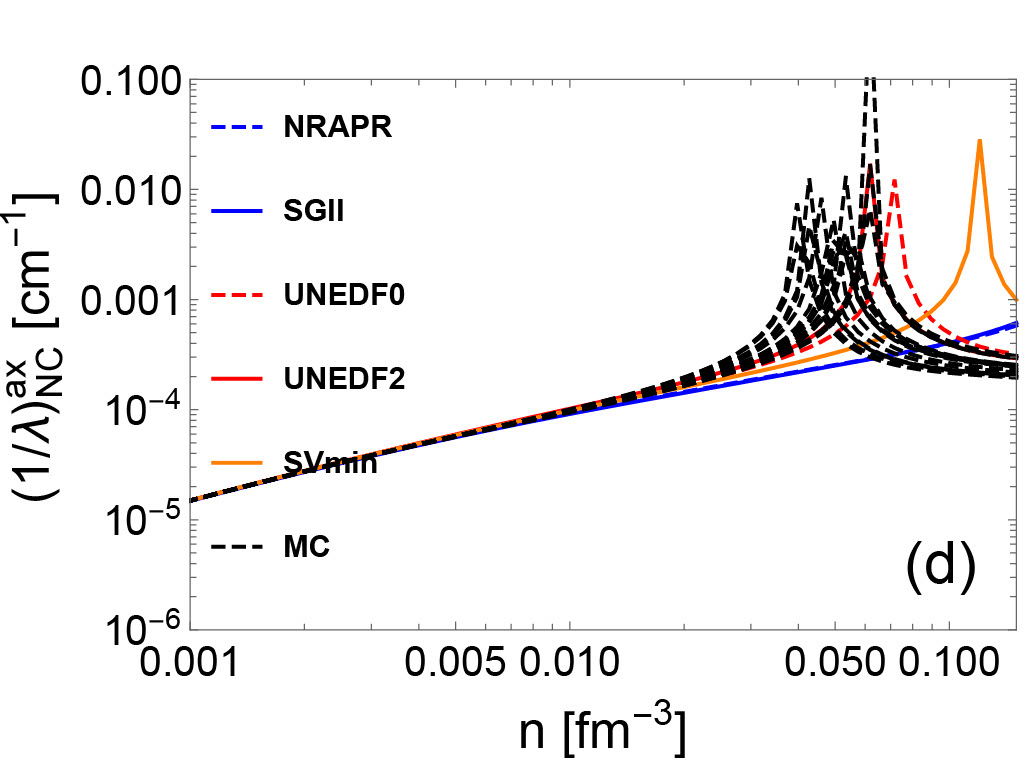}
         
	\caption{CC (panel a and b) and NC (panel c and d) IMFPs in vector current (panel a and c) and axial-vector current (panel b and d) channel at T=10 MeV. The colored solid curves stand for IMFPs derived from Skyrme-type interactions. The black dashed curves stand for IMFPs derived from MC EoSs, see section \ref{sec:formula} for detailed description of the underlying EoSs used to derive the residual interactions. } 
\label{fig:imfp}	
\end{figure*}

\begin{figure*}[htp]
	\centering
	\includegraphics[width=0.45\textwidth]{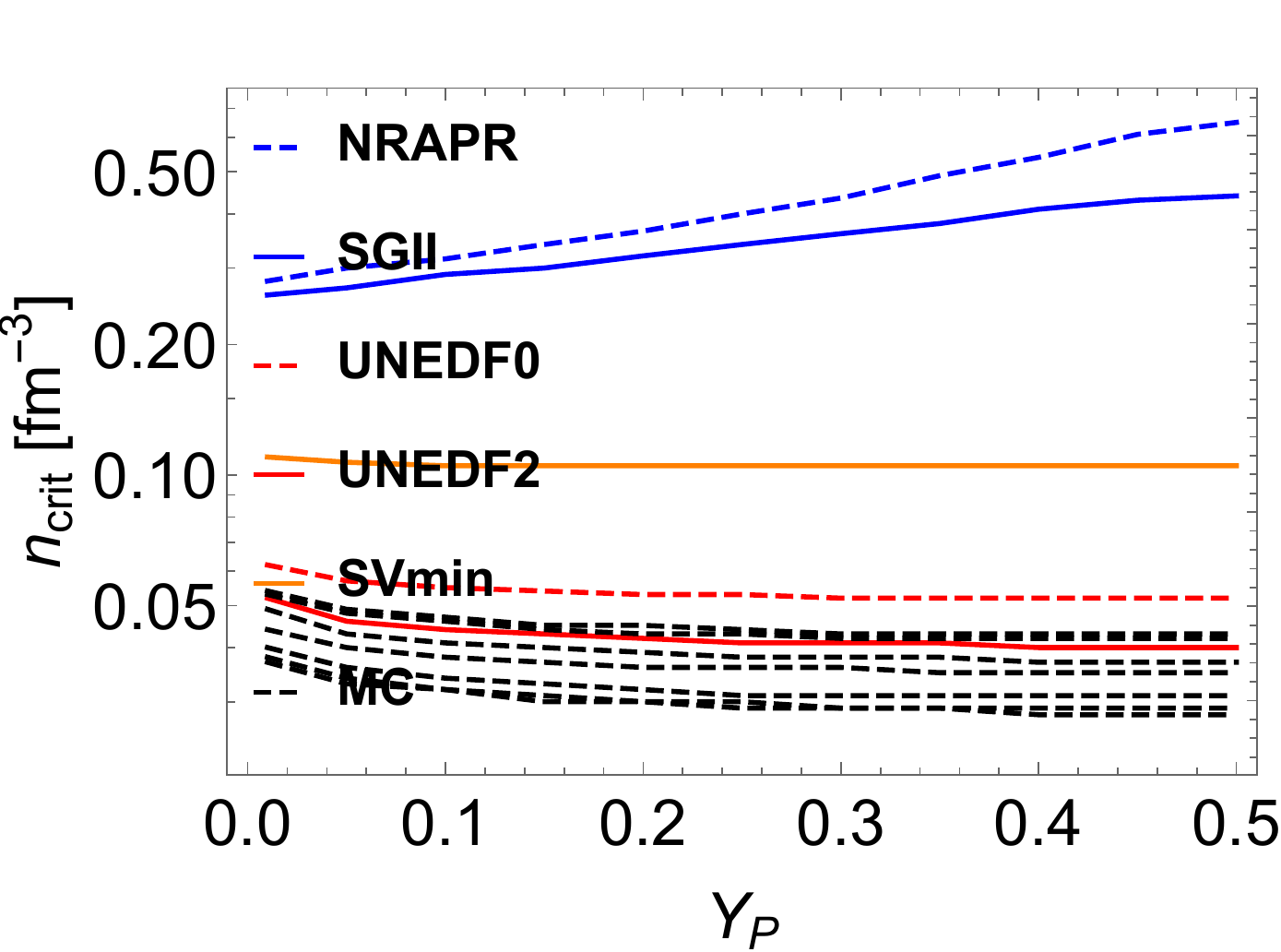}

	\caption{Ferromagnetic instability density $n_{crit}$ as a function of $Y_p$ in different EoSs. } 
\label{fig:spininstability}	
\end{figure*}

\begin{figure*}[htp]
	\centering
	\includegraphics[width=0.45\textwidth]{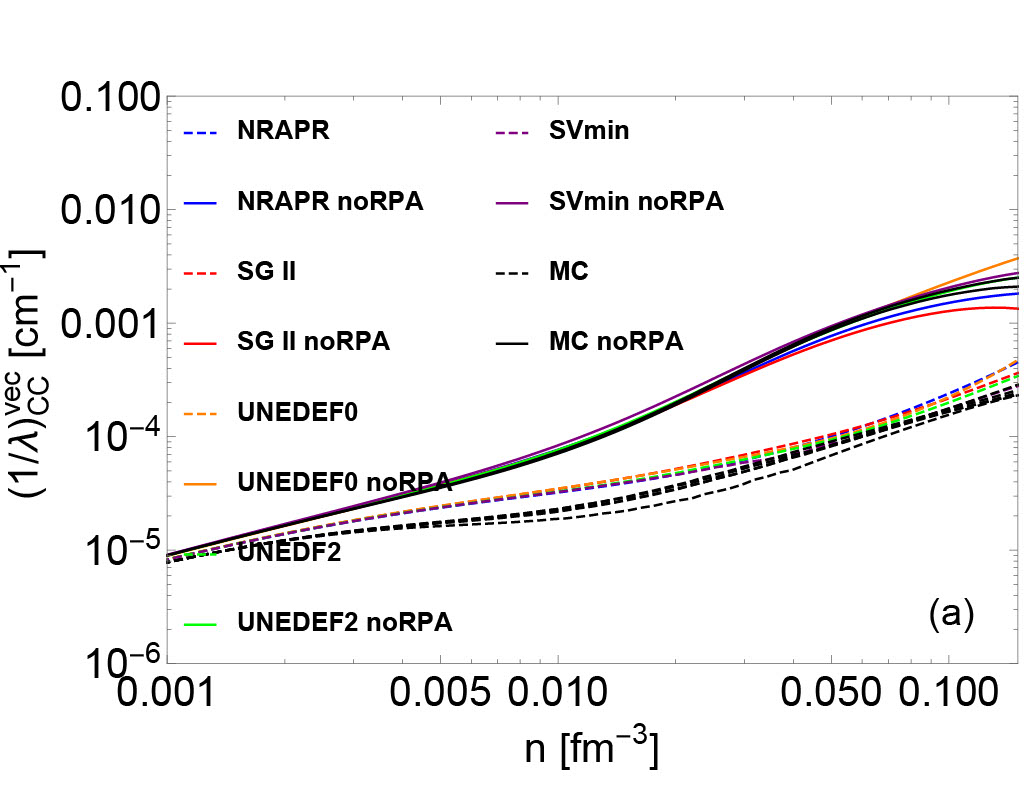}
    \includegraphics[width=0.45\textwidth]{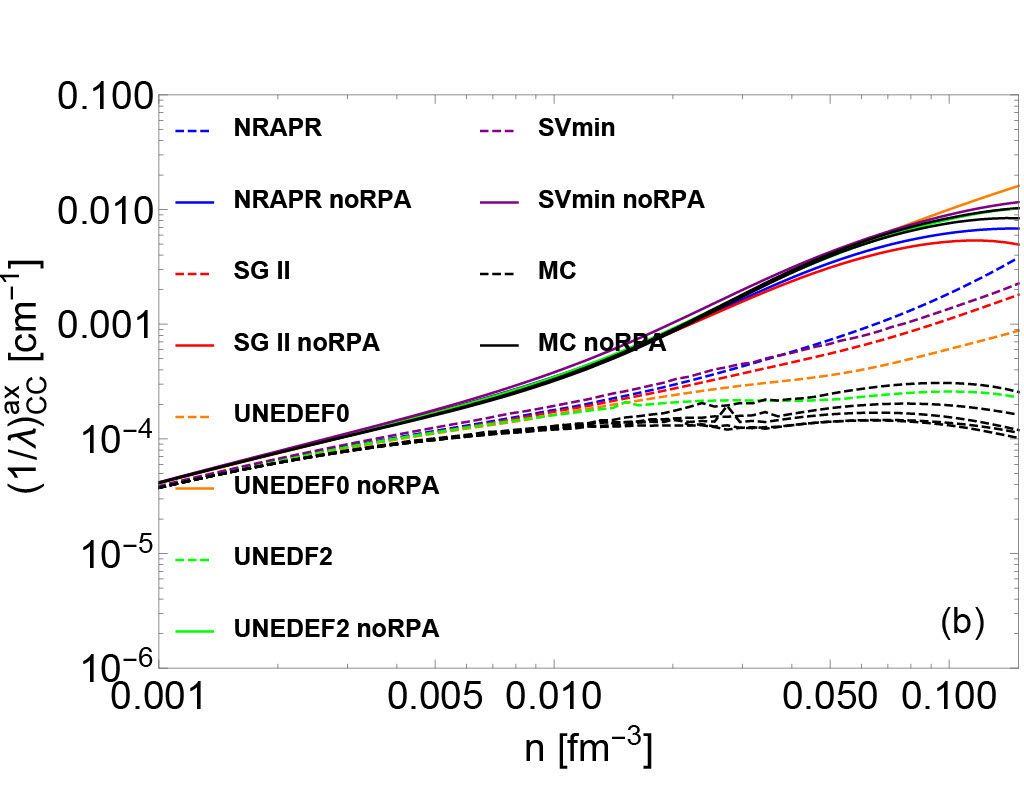}
    \includegraphics[width=0.45\textwidth]{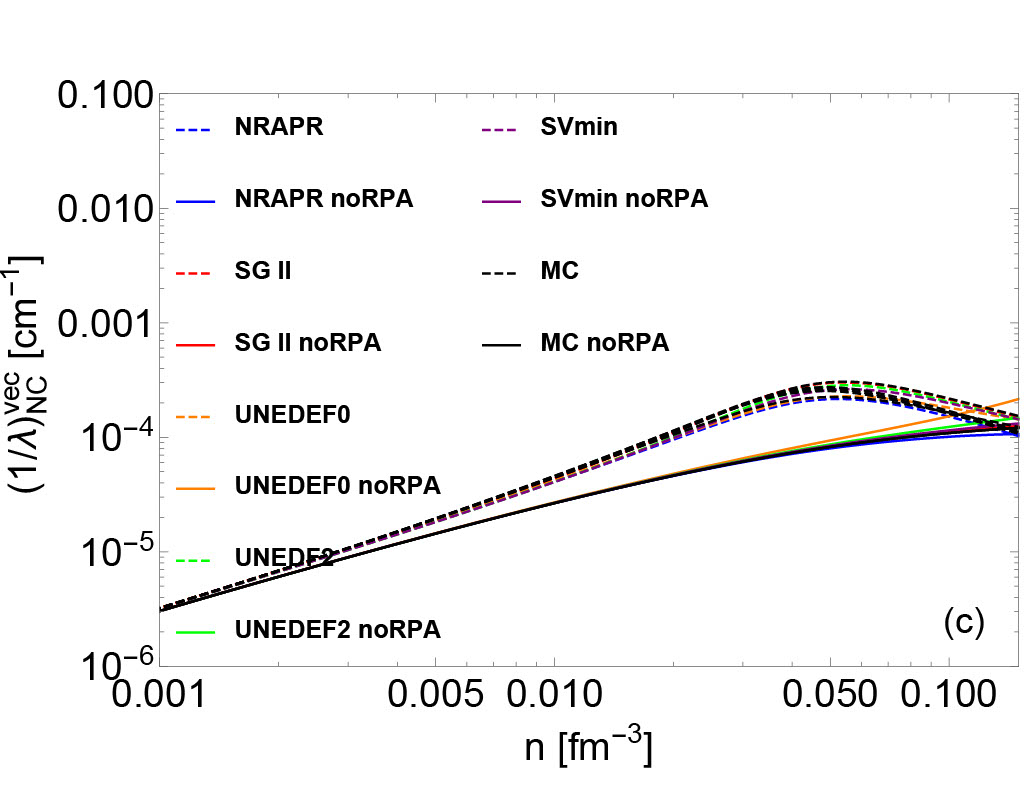}
    \includegraphics[width=0.45\textwidth]{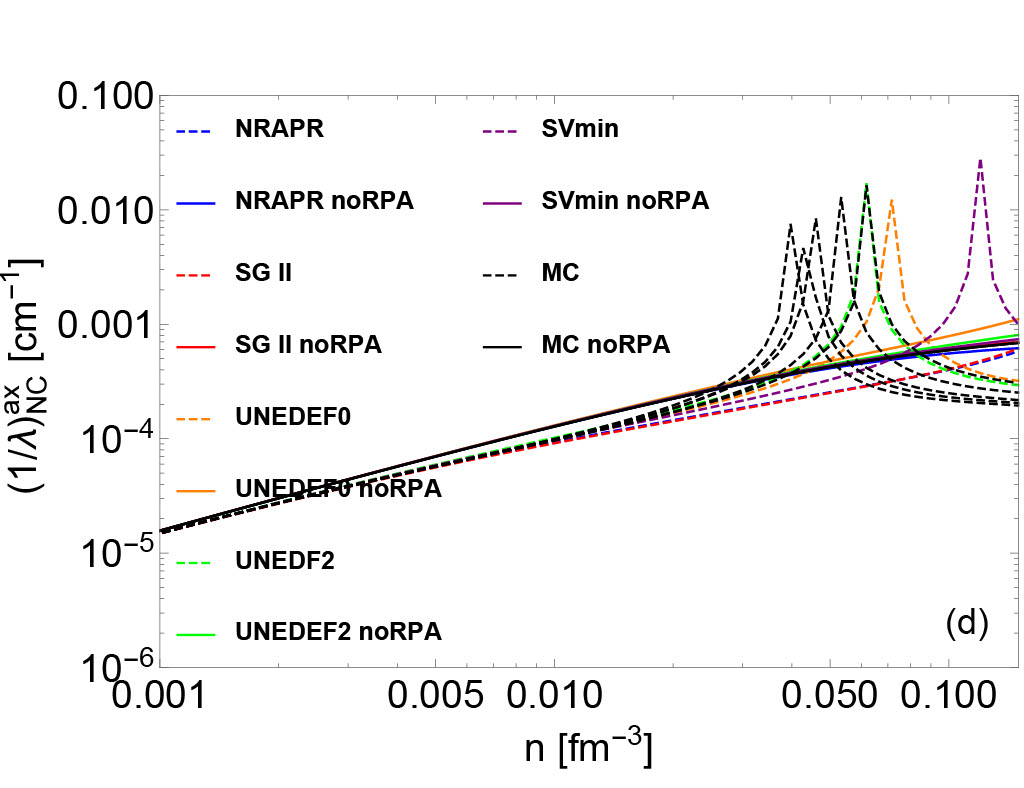}
	\caption{CC (panel a and b) and NC (panel c and d) IMFPs with and without RPA corrections in vector (panel a and c) and axial vector (panel b and d) channel at T=10 MeV. The colored solid curves stand for IMFPs derived from Skyrme and MC EoSs without RPA corrections. The colored dashed curves stand for IMFPs derived from Skyrme and MC EoSs with RPA corrections. The small peaks observed in a few MC curves are due to the numerical noise. } 
\label{fig:imfp_norpa}	
\end{figure*}

\begin{figure*}[htp]
	\centering
	\includegraphics[width=0.45\textwidth]{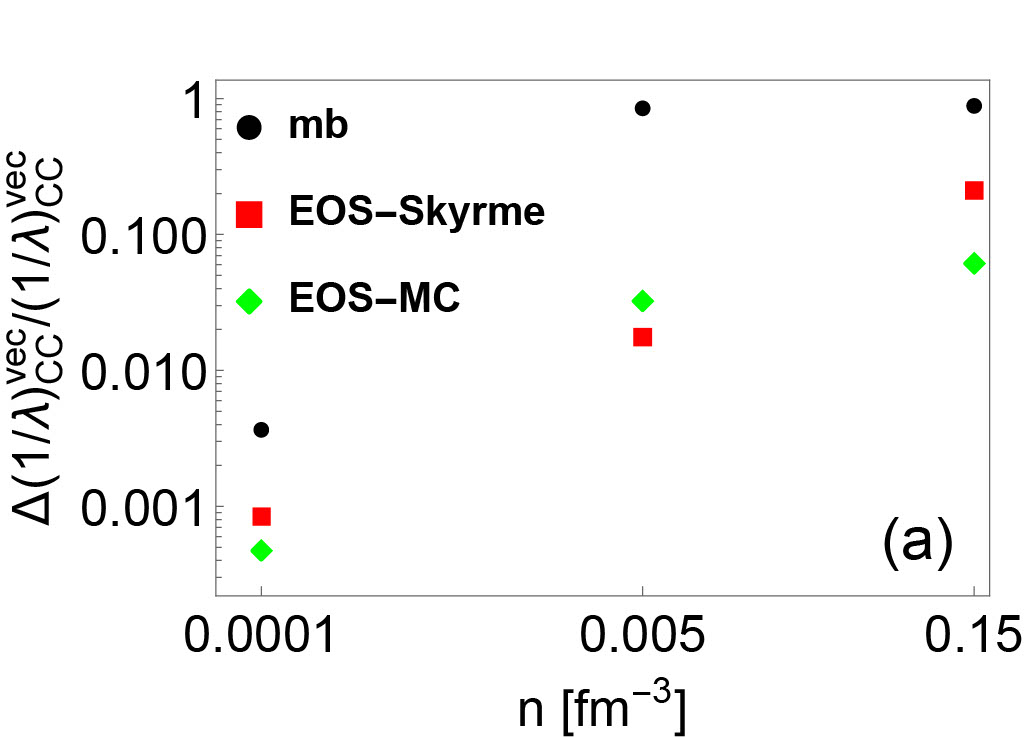}
	\includegraphics[width=0.45\textwidth]{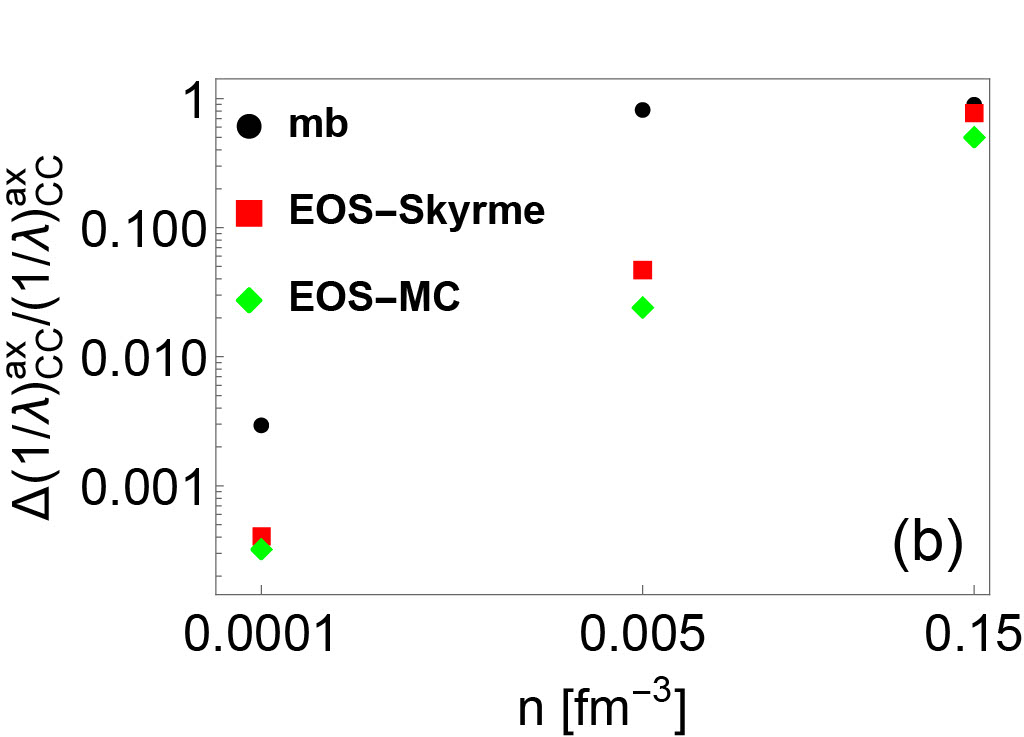}
	\includegraphics[width=0.45\textwidth]{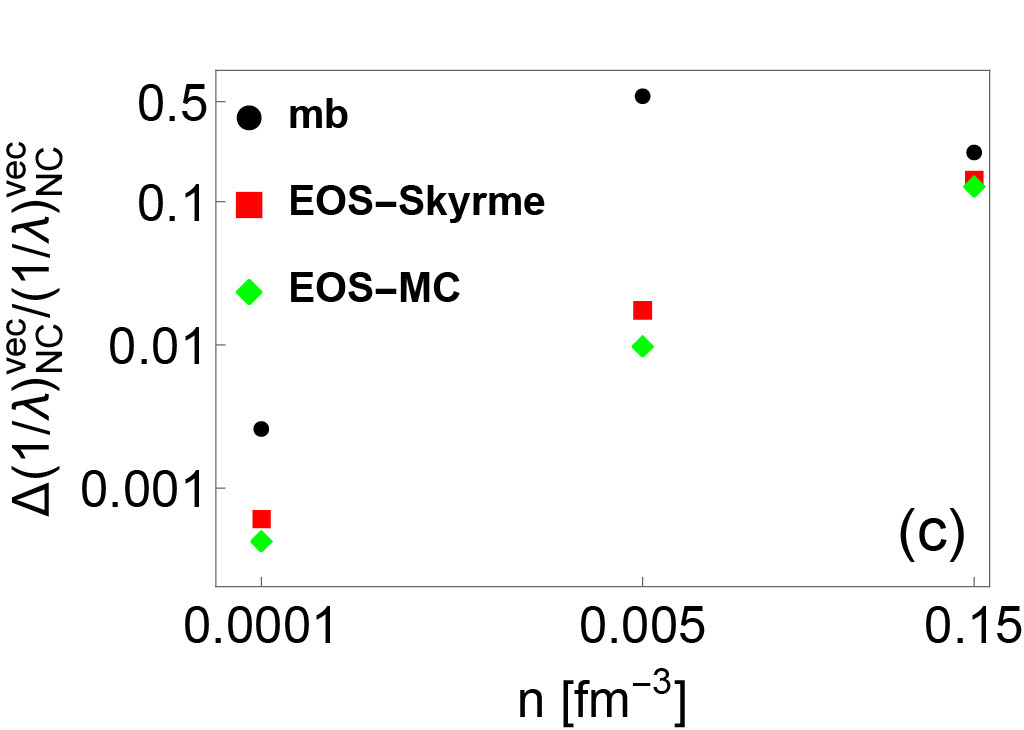}
	\includegraphics[width=0.45\textwidth]{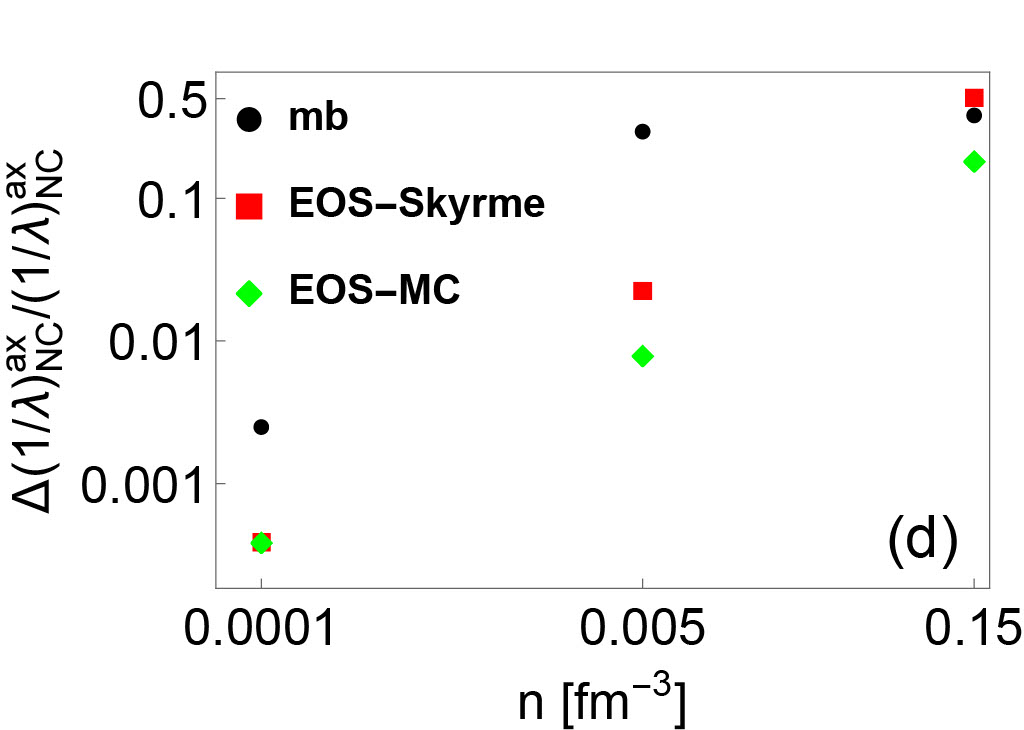}

	\caption{This figure shows the CC (panel a and b) and NC (panel c and d) uncertainties of IMFPs in vector (panel a and c) and axial vector (panel b and d). The points labeled ``mb'' show the uncertainty due to many-body methods by showing the difference between the RPA and mean-field estimates for pure Skyrme models. The points labeled ``EOS-Skyrme'' show the uncertainty due to the constraints used in obtaining the Skyrme-type EOSs. This was computed by comparing the variation across pure Skyrme models which were generated with fits to different sets of nuclear data. Finally, the points labeled ``EOS-MC'' show just the uncertainties in the mean free path across the Monte Carlo EOSs, which shows the variation due to the inability of the Skyrme model to precisely reproduce the experimental data from McDonnell et al. (2015). See section. \ref{sec:result} for detailed discussion of the classification of IMFP uncertainties.} 
\label{fig:imfpuncertainty}	
\end{figure*}

\begin{figure*}[htp]
	\centering
	\includegraphics[width=0.45\textwidth]{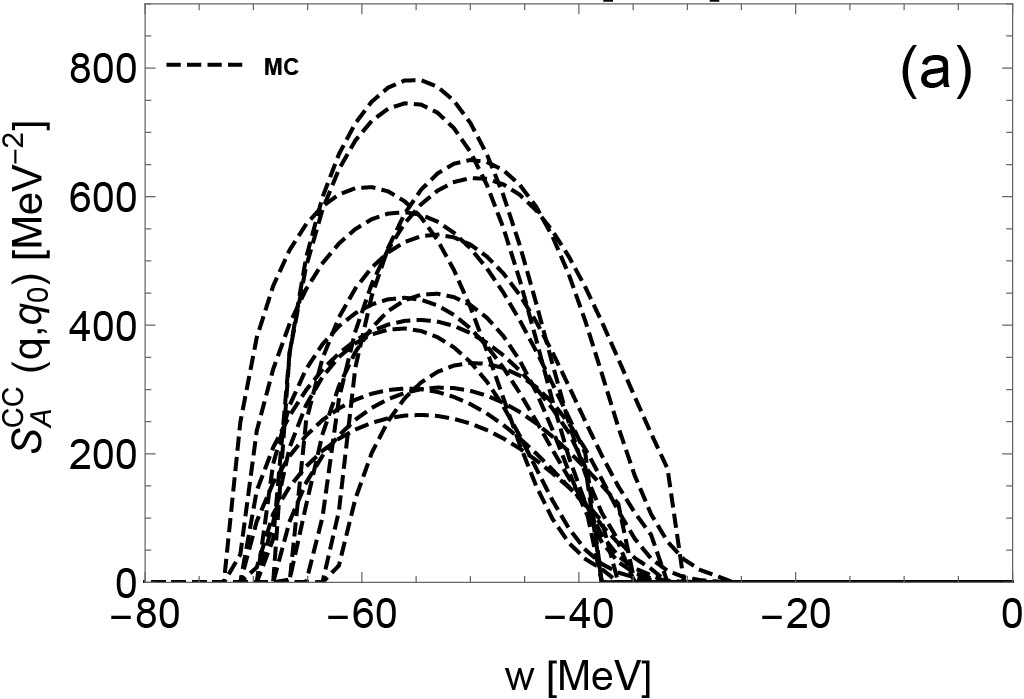}
	\includegraphics[width=0.45\textwidth]{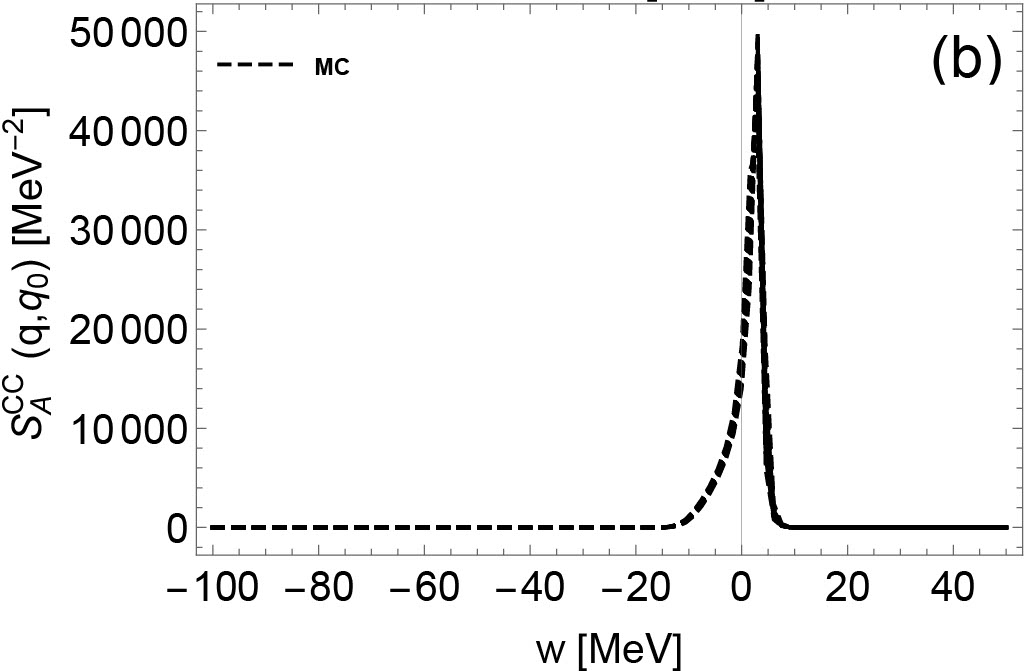}
	\includegraphics[width=0.45\textwidth]{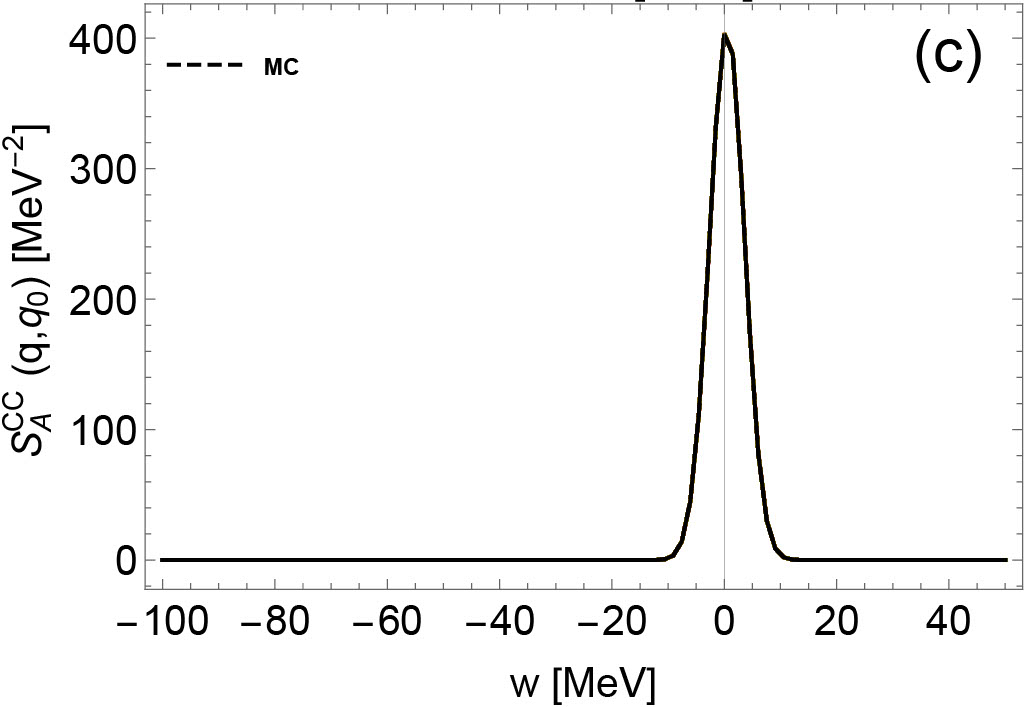}

	\caption{CC dynamic response based on MC EoSs in axial-vector current channel at T=10 MeV, with $q=3T=30~ \mathrm{MeV}$. The panel a, b and c are the dynamic response functions at $n=0.15$, $0.005$, $10^{-4}~\mathrm{fm^{-3}}$ respectively.} 
	\label{fig:ccdynamic}
\end{figure*}

\begin{figure*}[htp]
	\centering
	\includegraphics[width=0.45\textwidth]{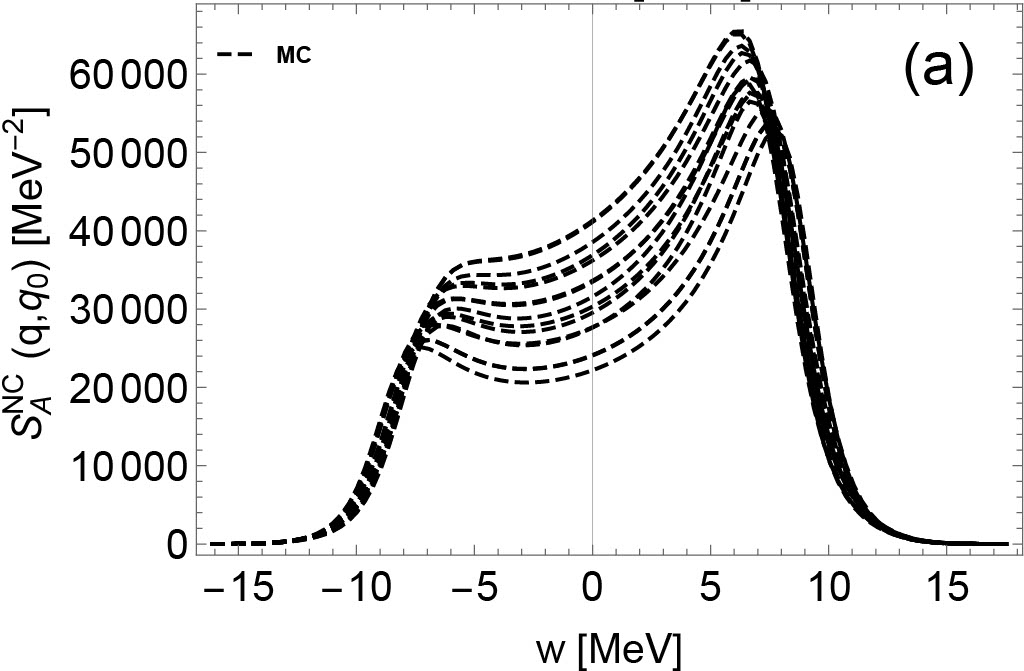}
	\includegraphics[width=0.45\textwidth]{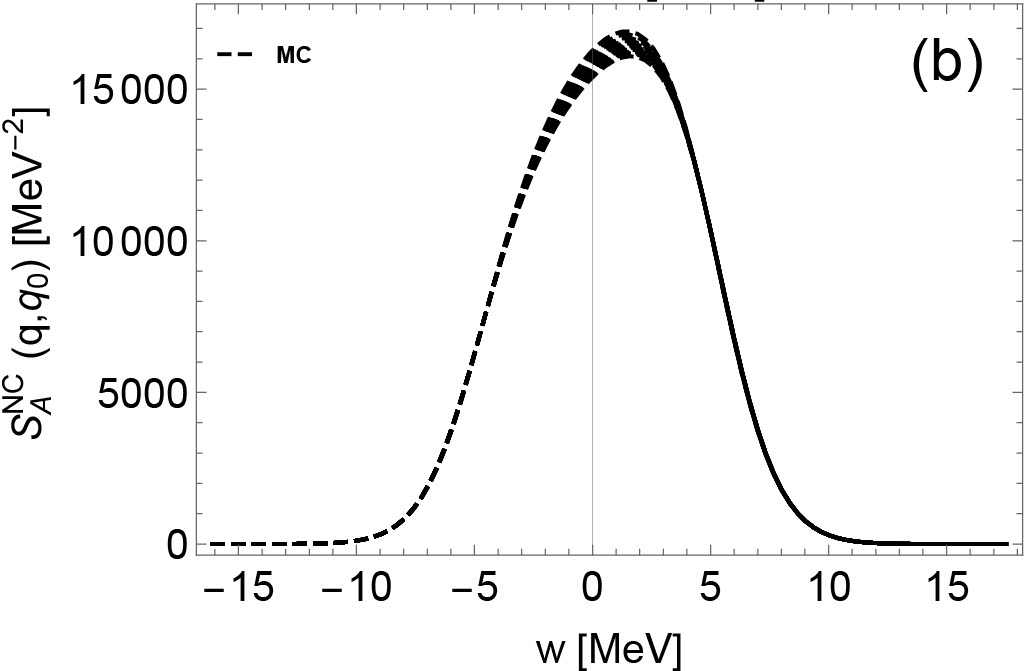}
	\includegraphics[width=0.45\textwidth]{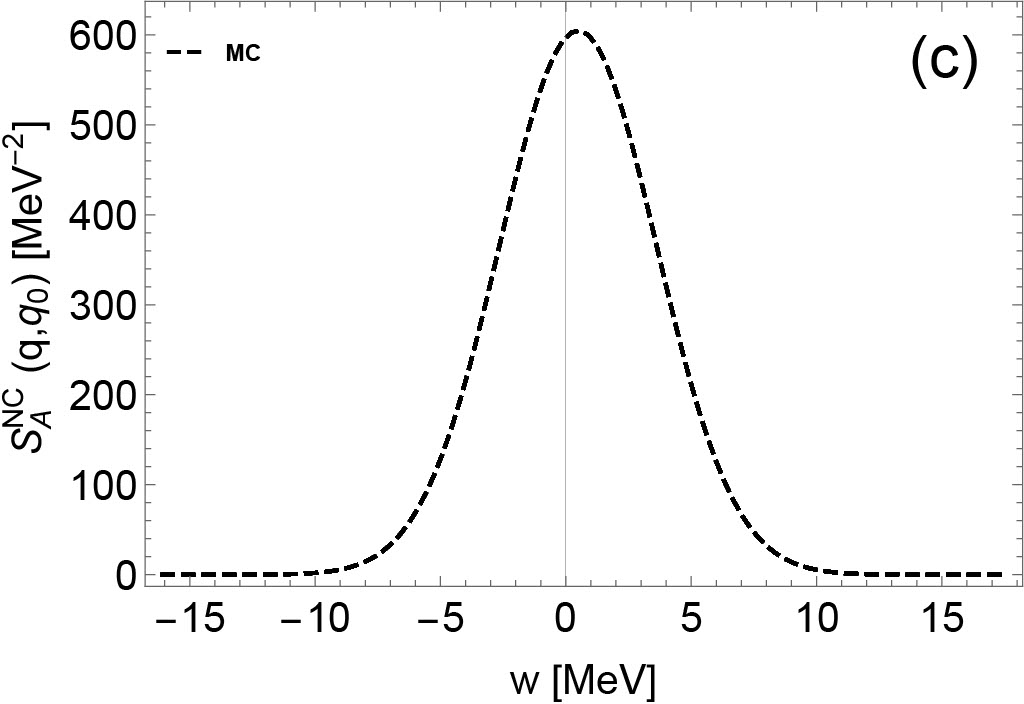}

	\caption{NC dynamic response based on MC EoSs in axial-vector current channel at T=10 MeV, with $q=3T=30~ \mathrm{MeV}$. The panel a, b and c are the dynamic response functions at $n=0.15$, $0.005$, $10^{-4}~\mathrm{fm^{-3}}$ respectively.} 
\label{fig:ncdynamic}	
\end{figure*}
 
 Concerning the EoS, we use a similar approach to that used in Ref.~\cite{PhysRevC.99.025803,Du22hd}. In particular, the benefit of this EoS is that it reproduces exactly the virial approximation in the nondegenerate limit, and in the high-density limit, it reproduces the Skyrme interaction. The low-density behavior is important for the neutrinosphere in core-collapse supernovae~\cite{Horowitz:2012us} where the Skyrme interaction fails to accurately describe the EoS. Our free energy is defined as
 \begin{eqnarray}
  f(n_B,x_p,T) &=& f_{\mathrm{virial}}(n_B,x_p,T) \eta \nonumber
  \\ && + f_{\mathrm{Sk}}(n_B,x_p,T) (1-\eta) \, .
\end{eqnarray}
where the function $\eta(z_n,z_p)$ is given by 
\begin{equation}
    \eta(z_n,z_p)=1/\big(1+10 (z_n^2+z_p^2) \big) \, ,
    \label{eq:gfunc}
\end{equation}
and $z_n$ and $z_p$ are the neutron and proton fugacities in the virial expansion. Note that, in Ref.~\cite{PhysRevC.99.025803}, 
slightly different coefficients were used in the denominator of $\eta$ but these have been modified to ensure the entropy is positive everywhere. The Skyrme free energy, $f_{\mathrm{Sk}}$, is taken from the UNEDF posterior distribution from Refs.~\cite{Kortelainen14,McDonnell15}.
The posterior distribution from Ref.~\cite{McDonnell15} takes the form of a table of 1000 Skyrme models, each selected from a likelihood function based on nuclear masses, charge radii, fision barriers, and other nuclear data. In Ref.~\cite{PhysRevC.99.025803} and in this work, we randomly select from that table of 1000 Skyrme models in order to obtain our results. The Skyrme parameters which we select are thus not uncorrelated, they retain the correlations which were obtained in Ref.~\cite{McDonnell15} as a result of the matching to the experimental data. These correlations are discussed below and displayed in Fig. 8.
In some cases, we
replace this Skyrme model with an alternate parameterization to test the variation beyond that obtained in this posterior. We use NRAPR~\cite{Steiner05ia} because it has been shown to be a good model for high-density matter~\cite{Dutra:2012mb}\ and is often used in EoS for core-collapse supernovae~\cite{Schneider:2019vdm}. We use SGII because it was explicitly constructed to fix the spin instability encountered in Skyrme models~\cite{VANGIAI1981379}. We also use the UNEDF0~\cite{Kortelainen10} and UNEDF2~\cite{Kortelainen14} EoS to compare with the original posterior distribution used in Refs.~\cite{Du22hd,PhysRevC.99.025803}. As in Ref.~\cite{PhysRevC.99.025803}, we also randomly select Skyrme models from the posterior distribution generated in Ref.~\cite{McDonnell15}. All of these models give a reasonable description of the binding energies, charge radii, and other experimental nuclear properties. The symmetry energy $S$ and its derivative $L$ are not taken from the posterior but used as additional parameters. Results obtained from EoS selected in this way are referred to as ``MC'' hereafter. In order to focus on the uncertainty of the neutrino opacities, we approximate the electron fraction in beta equilibrium to be the same for all models, $Y_e \approx 0.05+0.28 \exp[-126-31.49 \log_{10}(n_B)]$
 
 
 Given an EoS, the residual interactions, the nucleon chemical potentials and the nucleon effective mass were derived and then were applied in the neutrino opacity calculations. Note that, the effective mass is defined \emph{differently} in relativistic (the ``Dirac mass'') and in non-relativistic models (the ``Landau mass''). Consequently, one cannot directly use the non-relativistic type effective mass derived from Skyrme EoSs in relativistic neutrino opacity calculations.

 
 In the present study, the uncertainties in the neutrino opacities are directly resulting from the uncertainties in the EoS.
The aforementioned Skyrme-type EoSs are different to each other mainly because they are constrained by different experimental measurements/astronomical observations. In this way, we 
evaluate the impact on the neutrino opacities due to changing the Skyrme interaction, e.g., NRAPR, SGII, UNEDF0, UNEDF2 and SVmin.
Similarly, the impact of the MC EoSs on the neutrino opacities are also evaluated. Since the MC EoSs are constrained differently from the other Skyrme interactions previously mentioned, one could estimate the contribution of the nuclear and astrophysical uncertainties on the prediction of the neutrino opacities. This will be done in the discussion of our results in the following.

 \subsection{Correlations and Uncertainties}

In the following we discuss uncertainties of the neutrino IMFPs, in the framework of HF+RPA. As shown in Eqs.~\eqref{eq:CCrpaPIax}, \eqref{eq:CCrpaPIvec}, \eqref{eq:NCrpaPIax} and Eq.~\eqref{eq:NCrpaPIvec}, the input for the calculation of IMFPs based on HF+RPA are EoS-based quantities such as $M^*_\tau$, $M^*_{\tau'}$, $\mu_{\tau}$, $\mu_{\tau'}$, and $V_{ph}$. We first investigate the Pearson correlations between (1) two different EoS-based quantities; 
(2) EoS-based quantities and IMFPs and (3) two IMFPs at different densities.
The Pearson coefficient is given by: 
\begin{equation}\label{eq:pearson}
    \rho(A, B)=\dfrac{\sum_{M=1}^N (A_M-\Bar{A})(B_M-\Bar{B})}{\sqrt{\sum_{M=1}^N(A_M-\Bar{A})^2}\sqrt{\sum_{M=1}^N(B_M-\Bar{B})^2}},
\end{equation}
where $A$ and $B$ are (1) EoS-based quantities and EoS-based quantities; (2) EoS-based quantities and IMFPs and (3) IMFPs and IMFPs of a specific model $M$. In this work the EoS-based quantities are from $N$ hybrid EoSs, which were introduced with more details in Ref. \cite{Du22hd}. Given EoS-based quantities from $N$ models we define the covariance matrix $C_{ij}$ \cite{Fattoyev:2011ns}:
\begin{equation}\label{eq:cij}
    C_{ij}=\dfrac{1}{N}\sum_{M=1}^N x_{i,M}x_{j,M},
\end{equation}
where $x_{i,M}=(P_{i,M}-\Bar{P}_i)/\bar{P}_{i}$ and $P_{i,M}$ is the $i$th EoS-based quantity (e.g. nucleon effective mass, residual interaction,
etc)
predicted in EoS model $M$. 
The variance of IMFP is:
\begin{equation}\label{eq:sigmaimfp}
    \sigma_A^2=\sum_{i,j=1}^F\dfrac{\partial A}{\partial x_i}C_{ij}\dfrac{\partial A}{\partial x_j},
\end{equation}
where $A$ is the IMFP.

From Eqs.~\eqref{eq:cij} and \eqref{eq:sigmaimfp}, it is clear that the IMFP uncertainties are not only determined by the diagonal matrix elements of $C_{ij}$, but also by its off-diagonal matrix elements. Thus, the density-dependent correlations between EoS-based quantities may induce non-trivial density-dependent IMFPs uncertainties and non-trivial correlations between IMFPs in different channels.

\section{Results}\label{sec:result}

In this section we show our results for the IMFP, as well as the residual interactions which have been used in our study.

\subsection{Neutrino response from low-density to high-density region}

\begin{figure*}[htp]
	\centering
	\includegraphics[width=0.45\textwidth]{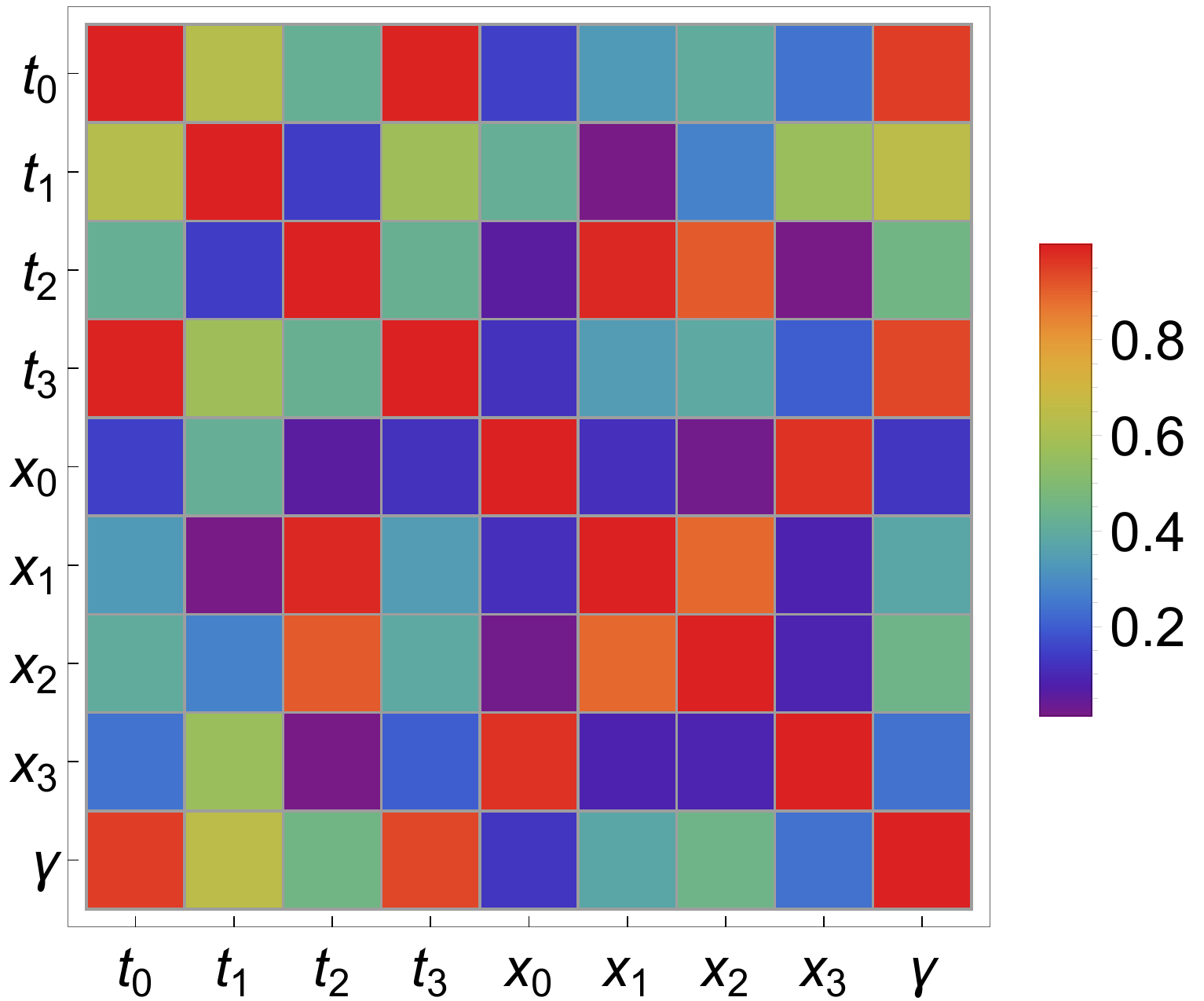}
	
	\caption{Pearson coefficients between Skyrme parameters $t_i$, $x_i$ and $\gamma$. } 
\label{fig:pearsonsksk}	
\end{figure*}

\begin{figure*}[htp]
	\centering
	\includegraphics[width=0.45\textwidth]{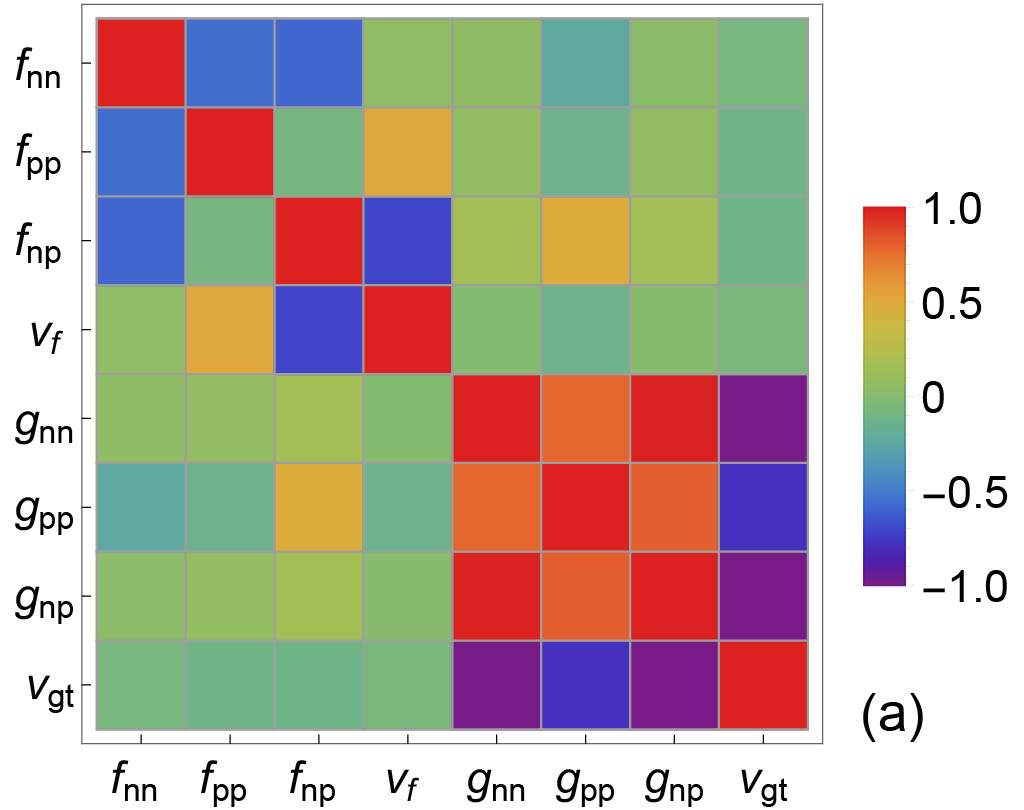}
	\includegraphics[width=0.45\textwidth]{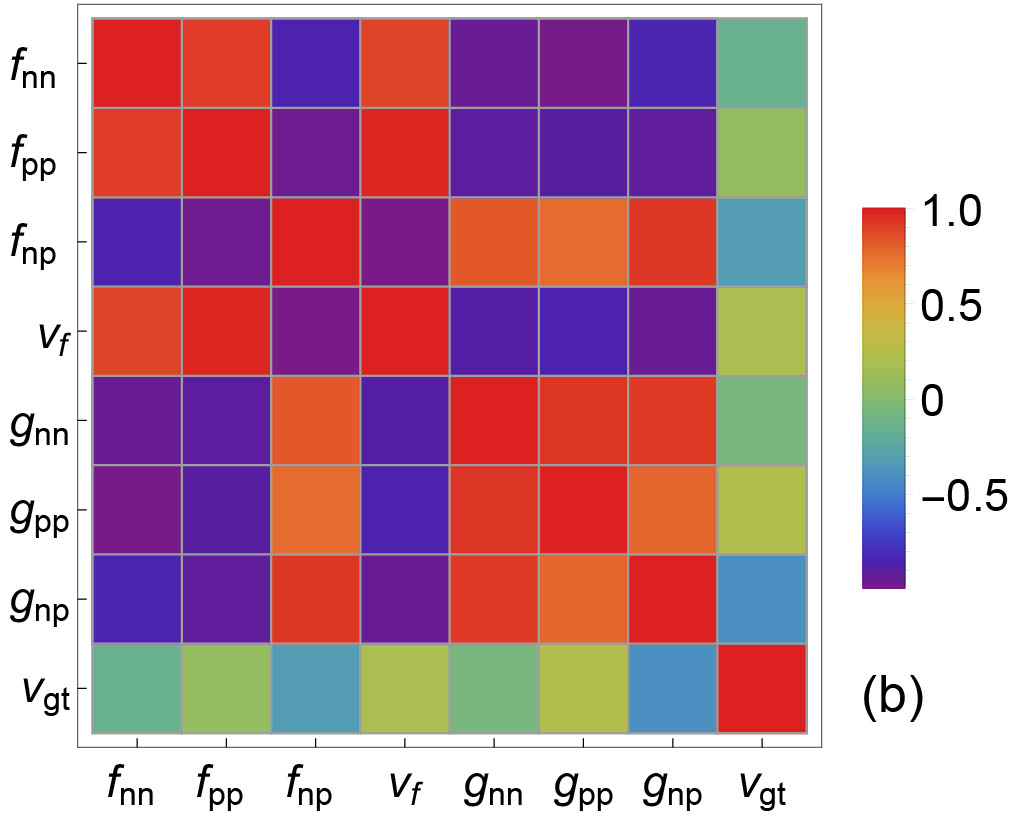}
	\includegraphics[width=0.45\textwidth]{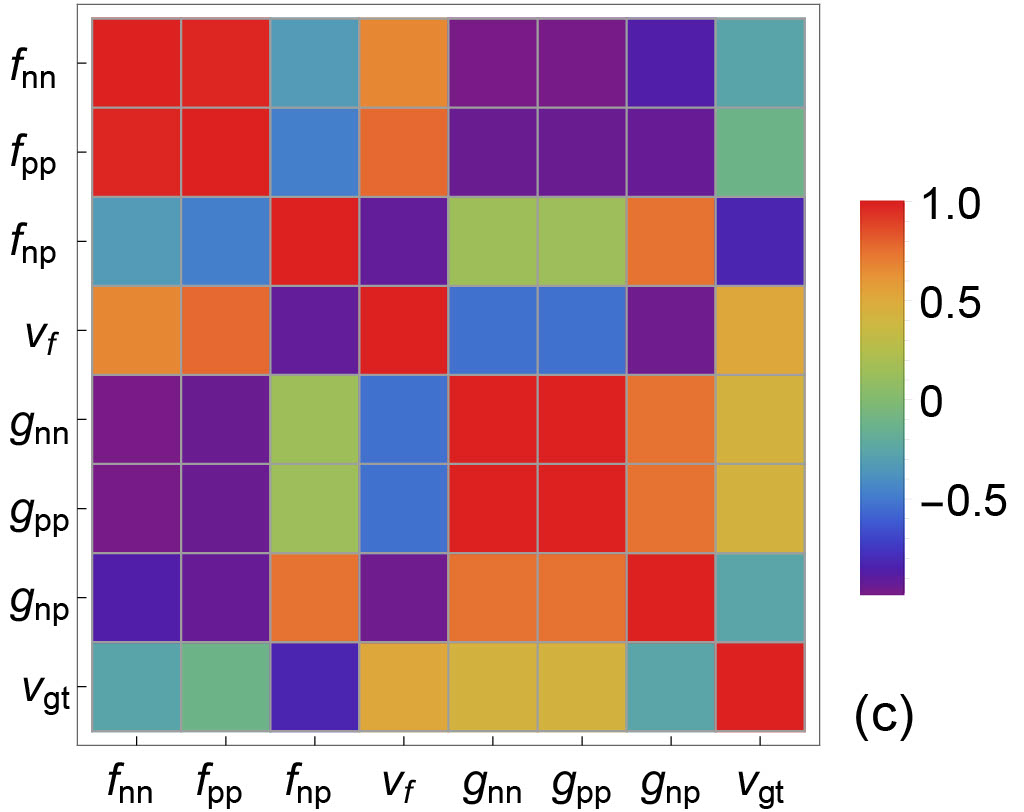}
	
	\caption{Pearson coefficients between residual interactions at $n=0.15~\mathrm{fm^{-3}}$ (a), $n=0.005~\mathrm{fm^{-3}}$ (b) and $n=10^{-4}~\mathrm{fm^{-3}}$ (c). } 
\label{fig:pearsonLFLF}	
\end{figure*}

\begin{figure*}[htp]
	\centering
	\includegraphics[width=0.45\textwidth]{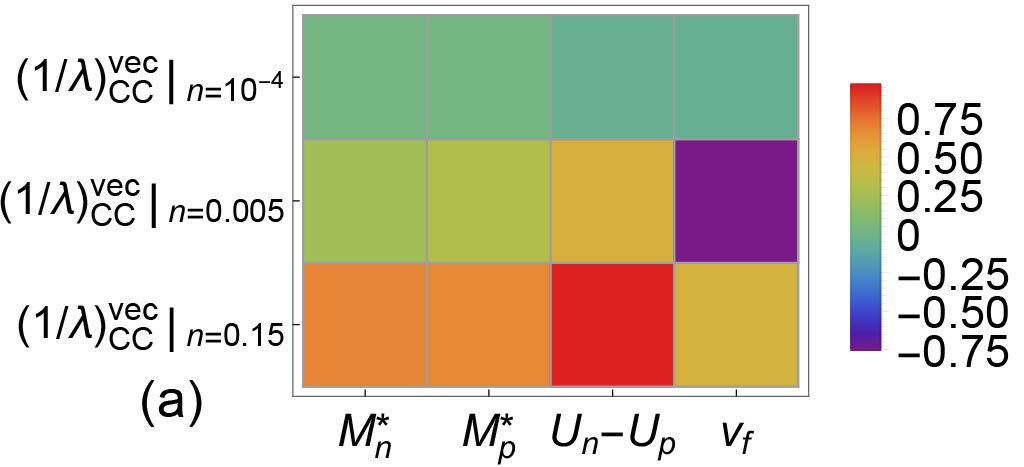}
	\includegraphics[width=0.45\textwidth]{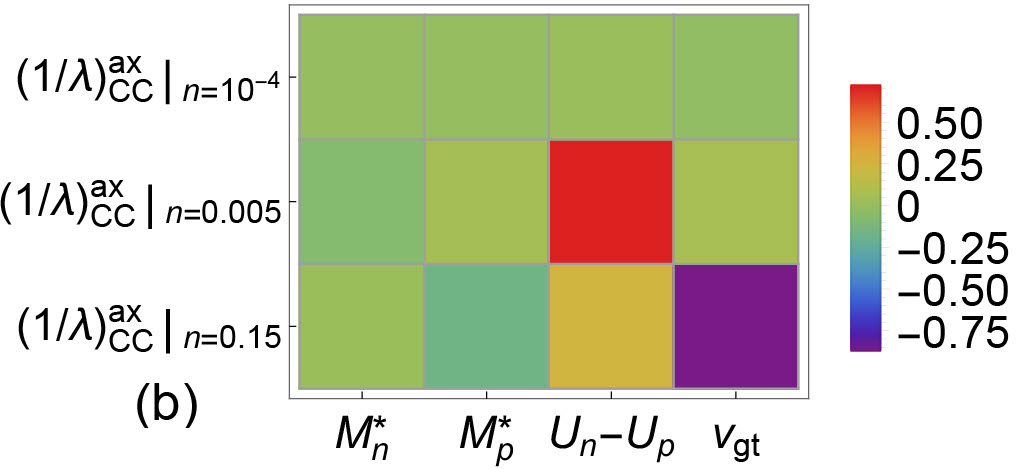}
	
	\caption{Pearson coefficients between EoS-based quantities and CC IMFPs in vector channel (a) and in axial vector channel (b). } 
\label{fig:pearsonLFIMFPcc}	
\end{figure*}

\begin{figure*}[htp]
	\centering
	\includegraphics[width=0.45\textwidth]{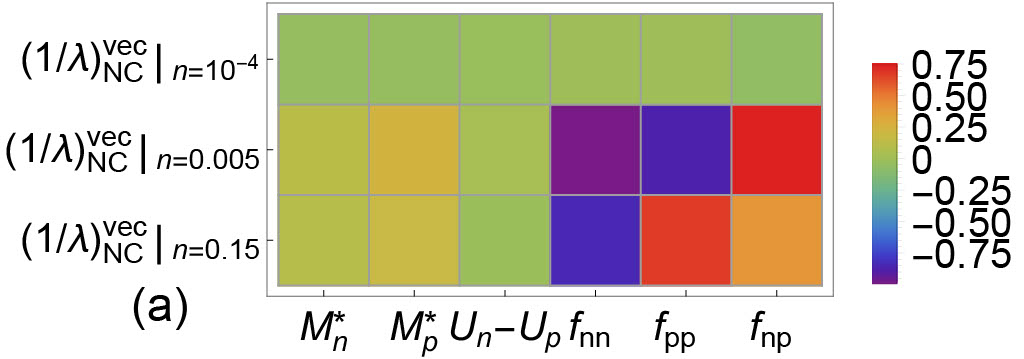}
	\includegraphics[width=0.45\textwidth]{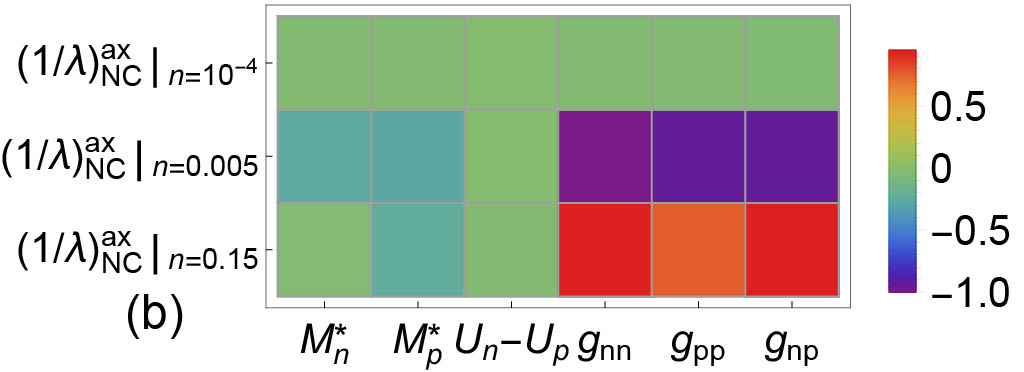}
	
	\caption{Pearson coefficients between EoS-based quantities and NC IMFPs in vector channel (a) and in axial vector channel (b). } 
\label{fig:pearsonLFIMFPnc}	
\end{figure*}

\begin{figure*}[htp]
	\centering
	\includegraphics[width=0.45\textwidth]{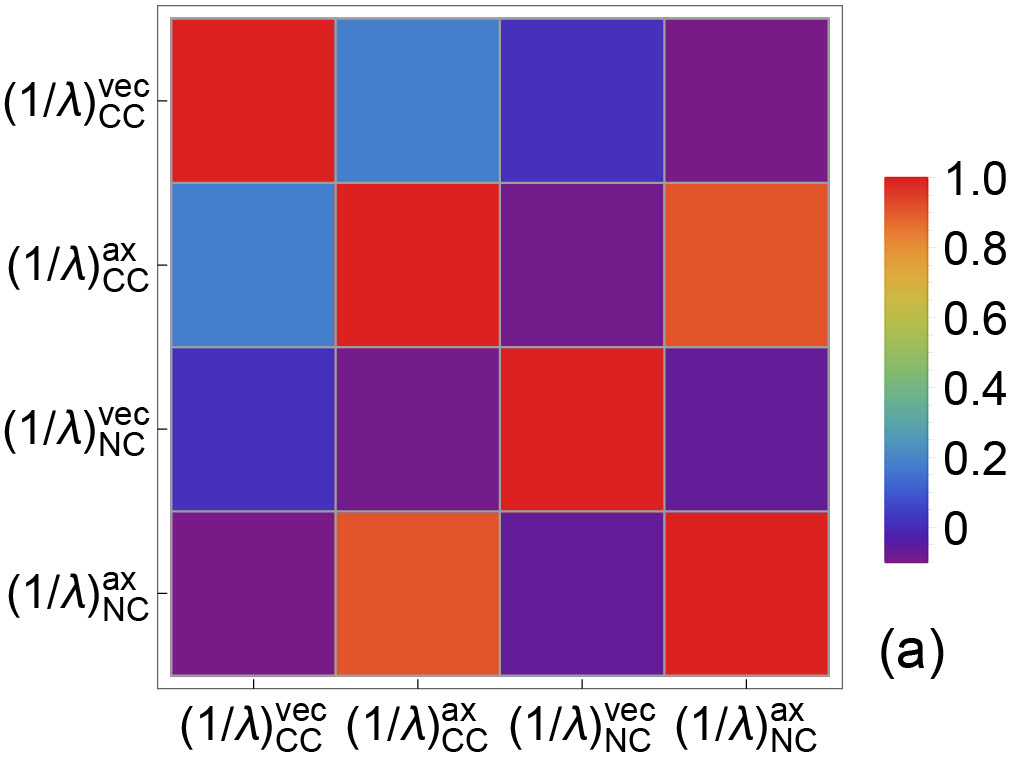}
	\includegraphics[width=0.45\textwidth]{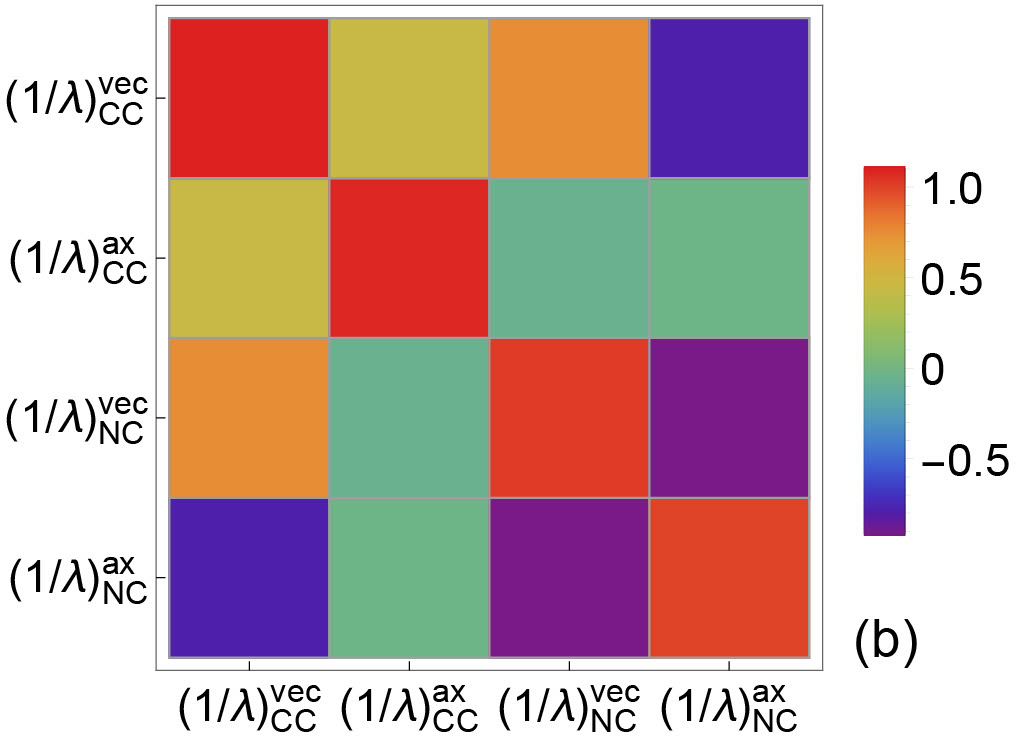}
	\includegraphics[width=0.45\textwidth]{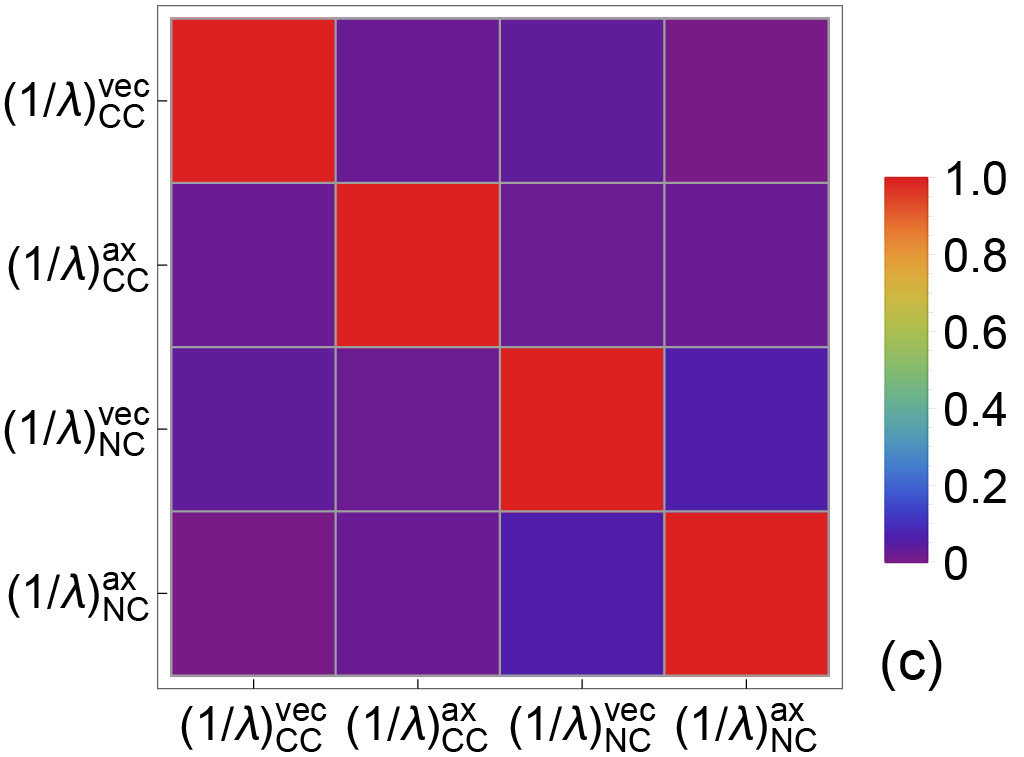}
	
	\caption{Pearson coefficients between IMFPs at $n=0.15~\mathrm{fm^{-3}}$ (a), $n=0.005~\mathrm{fm^{-3}}$ (b) and $n=10^{-4}~\mathrm{fm^{-3}}$ (c). } 
\label{fig:pearsonIMFPIMFP}	
\end{figure*}

In this subsection, we present the density-dependent residual interactions based on a set of EoSs previously introduced (the full derivation of EoS-based residual interactions is presented in Appendix~\ref{appendix1}), the IMFPs from low to high densities, and the dynamic responses at different densities. 

In Fig. \ref{fig:phinteractions}, the density-dependent residual interactions are shown. To illustrate the impact of EoSs on residual interactions, two groups of residual interactions are shown. The first group of residual interactions are based on various Skyrme models from low-density to high-density regime. The second group of residual interactions denoted as "MC" are Monte Carlo realizations based on the hybrid EoS model first introduced in \cite{PhysRevC.99.025803}, which reproduces the virial EoS in the low-density regime~\cite{Horowitz:2006pj,Horowitz:2016gul}. In the high-density regime, the residual interactions based on the hybrid EoSs sample the uncertainties in the parameter distribution originating from the UNEDF2 model. More details about the density-, temperature-, and isospin-dependent residual interactions are introduced in Appendix \ref{appendix1}. Note that we do not consider the uncertainties in the virial EoS resulting from the contribution from higher order virial coefficients since they are not well-constrained by experiments and observations. They contribute only as the density increases. At low densities,
the residual interactions from all of the MC EoSs converge because they are based on the same 2nd order virial coefficients. In our approach, the residual interactions in the low-density regime are constrained, for the first time, by a model-independent virial EoS. The density-dependent residual interactions beyond this low-density regime is however still poorly constrained by data. 

In the following, we briefly summarize the qualitative features of residual interactions in various channels. At around saturation density, CCSNe matter is very neutron-rich. In neutron-rich matter, the residual interactions $f_{nn}$ and $g_{nn}$ play a major role in NC interactions, while the influence from $f_{pp}$ (which is dominated by the Coulomb interaction) is weak. Indeed, in pure neutron matter, $f_{nn}=f_0$ and $g_{nn}=g_0$. As densities increase, $f_{nn}$ increases and becomes positive, which is a feature expected on general grounds and originating from the dominant contribution of the repulsive vector meson.
The density dependence of $g_{nn}$ is not well constrained by nuclear experiments probing collective modes, even at saturation density. In Skyrme models, we observe that $g_{nn}$ decreases as the density increases, and may become negative at high densities. In CC reactions, $V_\mathrm{f}$ and $V_\mathrm{gt}$ are residual interactions relevant to vector and axial vector neutrino-nucleon reactions. These residual interactions are calculated consistently from low-density to high-density regimes, and their detailed derivation are shown in the Appendix (see Eqs. ~\eqref{eq:vfgeneral} and ~\eqref{eq:vgtgeneral}). 
All the Skyrme models that we have tested in our study predict consistent values for $V_\mathrm{f}$, which decreases as the density increases.
In the axial channel however, the dispersion among the Skyrme predictions is larger than in the vector channel.
As shown in the upper left panel of Fig. \ref{fig:phinteractions}, the values of $V_\mathrm{gt}$ based on Skyrme models diverge in high-density regime. Overall, we observe that the uncertainties of residual interactions in spin-dependent channels are obviously larger than the uncertainties in spin-independent channels, which reflects the lack of experimental constraints for the time-odd terms in Skyrme EDFs and thus the phenomenological spin-dependent forces.  

In Fig. \ref{fig:imfp}, vector (left panels) and axial vector (right panels) IMFPs of neutrino-nucleon reactions are plotted. Since the calculation of neutrino opacities are consistent with the underlying EoSs, we observe that the uncertainties of EoS-based quantities (e.g. residual interactions/effective mass/nucleon chemical potentials) result in variations of IMFPs. In high-density regime, the uncertainties of axial vector IMFPs in both CC (upper right panel) and NC (lower right panel) are larger than the uncertainties of vector IMFPs (upper left and lower left panels)
. This is because the axial vector IMFPs are sensitive to the residual interactions $V_\mathrm{gt}$, $g_\mathrm{nn}$, $g_\mathrm{np}$ and $g_\mathrm{pp}$, which are derived from the poorly-constrained time-odd part of the Skyrme EDF. 

In Eqs.~ \eqref{eq:CCrpaPIvec}, ~\eqref{eq:CCrpaPIax}, ~\eqref{eq:ncvecPI} and ~\eqref{eq:ncaxPI}, we observe that the polarization functions containing the real parts have poles, and these poles may result in collective excitations \cite{GarciaRecio1992,Burrows:1998cg}. For both the vector and the axial vector channels, the collective excitations in the medium appear if the reaction has a $(q, q_0)$ pair that satisfies the mode’s dispersion relation. The resonances may
enhance
the IMFPs for the neutrino-nucleon reactions, similarly to the effect of Giant-dipole resonance and the Gamow-Teller resonance on finite nuclei scattering. As shown in the right lower panel of Fig. \ref{fig:imfp}, a significant increase is observed in those IMFPs based on UNEDF and MC EoSs in high-density regime. In this region, the $g_{nn}$ values decrease and become negative, representing an attractive potential enhancing the IMFP. The increase of IMFPs observed in axial NC panel may be due to the Gamow-Teller collective mode. However, we stress that the collective modes are sensitive to the strength of the related residual interactions which are not well-constrained, and may result in the creation of a ferromagnetic unstable region. There are no clear experimental observations to support the existence of ferromagnetic instability in neutron stars and CCSNe. We show in Fig. \ref{fig:imfp} that if the ferromagnetic unstable region appears at (sub)saturation density, the NC neutrino opacity increases by several orders and such an increase may be associated with observable modulations on late-time neutrino signals from a future galactic CCSNe. We include the NC axial IMFPs with high peaks at high densities in Fig. \ref{fig:imfp} to maintain the consistency between the underlying EoSs and the neutrino IMFPs. In future work, we will construct new EoSs which allow us to quantify the uncertainties near the saturation density without a spin instability while still matching laboratory experiments. 

The ferromagnetic instability could be obtained from the long wave-length limit of the HF+RPA response. In this limit, the density at which matter undergoes a ferromagnetic instability is denoted $n_\mathrm{crit}$, which is defined from the zero of the matrix:
$$\begin{pmatrix}
  G_{nn}(n,Y_p)+1 & G_{np}(n,Y_p)\\ 
  G_{pn}(n,Y_p) & G_{pp}(n,Y_p)+1
\end{pmatrix}=0,$$ where $G_{nn}$, $G_{pp}$, $G_{np}$ and $G_{pn}$ are dimensionless Landau parameters. The values for $n_\mathrm{crit}$ as function of the proton fraction $Y_p$ is shown in Fig. \ref{fig:spininstability} for different EoSs. The ferromagnetic unstable region in UNEDF0 and UNEDF2 extends from $n\approx~0.05$~fm$^{-3}$ to higher densities. The UNEDF EoSs were constructed to describe the bulk ground state and collective excited state properties of various spherical atomic nuclei. However, in most atomic nuclei, the ground state and collective excited state properties are not sensitive to the \emph{spin-dependent} interactions. Consequently, the spin-dependent residual interactions in the UNEDF appraoch may be poorly constrained.  

In Fig. \ref{fig:imfp_norpa}, the IMFPs with and without the RPA correlations are shown in four different channels (CC vector/CC axial/NC vector/NC axial). In CC vector and CC axial channel, the RPA correlations 
lead to a decrease of the neutrino opacity.
In NC axial channel, the RPA correlations 
reduce (increase) the neutrino opacity in the high-density regime for NRAPR and SGII (UNEDF) EoSs, where the increase is due to the collective modes.
However, as mentioned above, the spin-dependent residual interactions in UNEDF may be poorly constrained and the RPA-amplified IMFPs may reside in ferromagnetic unstable region. In NC vector channel, the many-body effect manifested by the RPA correlations increase the neutrino opacity, which quantitatively agrees with the results of NC vector neutrino-nucleon scattering opacity based on model-independent virial expansion. Indeed, as shown in Fig. \ref{fig:phinteractions}, the residual interactions $f_{nn}$ that mainly controls the behavior of the RPA correlations in the NC vector channel were derived density-dependently, and in the low-density regime $f_{nn}$ reproduces the virial predictions.

By comparing IMFPs with and without RPA correlations, the variations of neutrino opacity resulting from two different basic many-body methods (HF and HF+RPA) can be evaluated. The variation between the HF and the HF+RPA predictions is
denoted as $\Delta(1/\lambda)_\mathrm{mb}$ in the following. By comparing IMFPs with RPA corrections but with different underlying EoSs, the variations of neutrino opacity resulting from the choice of Skyrme EoS constraints can be evaluated. This type of variation will be denoted $\Delta(1/\lambda)_{\mathrm{EOS-Skyrme}}$ in the following. By comparing IMFPs consistent with MC EoSs, we estimate the variations of neutrino opacity due to the uncertainties across the Monte Carlo EOSs, which shows the variation due to the inability of the Skyrme model to precisely reproduce the experimental data. This type of variation will be denoted as $\Delta(1/\lambda)_{\mathrm{EOS-MC}}$ in the following. Based on Fig. \ref{fig:imfp} and \ref{fig:imfp_norpa}, an estimate for the \emph{hierarchy} of these three types of neutrino opacity variations can be constructed in a wide range of densities. We find that approximately, $\Delta(1/\lambda)_{\mathrm{EOS-MC}}\le\Delta1/(\lambda)_{\mathrm{EOS-Skyrme}}<\Delta(1/\lambda)_\mathrm{mb}$. However, in axial vector channel, where the spin-dependent residual interactions are poorly-constrained, the $\Delta(1/\lambda)_{\mathrm{EOS-MC}}$ may be comparable with or even larger than the $\Delta\lambda_\mathrm{mb}$. The hierarchy of relative neutrino opacity uncertainties as a function of density are shown in Fig. \ref{fig:imfpuncertainty}.

In Figs. \ref{fig:ccdynamic} and \ref{fig:ncdynamic}, the dynamic responses are plotted for various densities. Since for most regions of the phase space the axial response dominates the neutrino-nucleon IMFPs, only the $S_A(q,q_0)^{CC}$ and $S_A(q,q_0)^{NC}$ are shown here. As density increases, we observe that the transferred energy $q_0$, favored by CC dynamic responses, deviates from $q_0=0~\mathrm{MeV}$ and become negative, due to the modifications from nucleon potential shifts in CC reactions \cite{Reddy:1997yr}. For NC dynamic response functions, the difference between neutron and proton single-particle energies do not have any influence on these reactions that conserve nucleon isospins. Consequently, the $S^{NC}(q,q_0)$ centers at $q_0\approx0~\mathrm{MeV}$ at any densities. Finally, we observe that the uncertainties in the dynamical response functions increase as density increases, due to the increasing uncertainties of EoS-based quantities (e.g. the residual interactions, the nucleon effective mass, and nucleon single-particle energies).

\subsection{Correlations between EoSs and Neutrino Response}

In this section we present the Pearson correlations $\rho$ between (1) the EoS-based quantities and the EoS-based quantities; (2) the EoS-based quantities and the IMFPs; (3) the IMFPs and the IMFPs, using Eq.~\eqref{eq:pearson}. The Pearson correlation coefficients quantitatively describes the connection between two quantities, and
allows us to estimate the degrees of correlations between different quantities, whether they may be or not be measured or observed.
In particular, a value of $\rho(A,B) = \pm1$ implies that the two observables are fully correlated/anticorrelated, whereas a value of $\rho(A,B) = 0$ means that the observables are totally uncorrelated. 

In Fig. \ref{fig:pearsonsksk}, the correlation coefficients 
among Skyrme parameters
are plotted. In Fig. \ref{fig:pearsonLFLF}, the correlation coefficients 
among the residual interactions 
are plotted for $n=10^{-4}$, $0.005$, $0.15~\mathrm{fm^{-3}}$. 
Only off-diagonal correlation are of interest in these two figures.
Note that several residual interactions in off-diagonal blocks are highly correlated/anti-correlated. For example, at $n=10^{-4}~ \mathrm{fm^{-3}}$ we observe that $\rho(f_{nn},f_{pp})\approx1$ and $\rho(g_{nn},g_{pp})\approx1$. Indeed, in low-density regime where the residual interactions are calculated based on the virial approximation, we have $f_{nn}=f_{pp}$ and $g_{nn}=g_{pp}$. Interestingly, the correlation coefficients between the residual interactions are density dependent. As shown in Fig. \ref{fig:pearsonLFLF}, as density increases $\rho(f_{nn},f_{pp})$ decreases. 
For the three different densities explored in Fig. \ref{fig:pearsonLFLF}, the correlations among $g_{nn}$, $g_{np}$ and $g_{pp}$ remains high. The density dependence of correlations between EoS-based quantities may result in (1) non-trivial correlations between IMFPs and EoS-based quantities; and (2) non-trivial correlations between IMFPs and IMFPs. 

In Fig.\ref{fig:pearsonLFIMFPcc} and Fig.\ref{fig:pearsonLFIMFPnc}, the correlations between EoS-based quantities and IMFPs are plotted. Correlations between IMFPs and EoS-based quantities reveal the sensitivity of a particular type of IMFP to EoS-based quantities. In Fig. \ref{fig:pearsonLFIMFPcc}, the correlations between the EoS-based quantities and the CC IMFPs in both vector and axial vector channels are shown. We observe that (1) the sensitivity of $M^*_\mathrm{n}$ ($M^*_\mathrm{p}$) to CC vector IMFPs are higher than that to CC axial vector IMFPs; (2) the density dependence of sensitivity of $U_n-U_p$ to CC IMFPs is different in vector and in axial vector channels; and (3) the density dependence of sensitivity of residual interactions to CC IMFPs is different in vector and in axial vector channels. In Fig. \ref{fig:pearsonLFIMFPnc}, the correlations between the EoS-based quantities and the NC IMFPs in both vector and axial vector channels are shown. We observe that (1) the sensitivity of $M^*_\mathrm{n}$ ($M^*_\mathrm{p}$) to NC IMFPs is moderate in both vector and axial vector channels; (2) the sensitivity of $U_n-U_p$ to NC IMFPs $\approx~0$, reflecting the fact that NC IMFPs are not modified by nucleon potential shifts; and (3) the density dependence of sensitivity of residual interactions to NC IMFPs is different in vector and in axial vector channels. 

In Fig. \ref{fig:pearsonIMFPIMFP}, the correlations between IMFPs and IMFPs are plotted. A strong correlation between a theoretically (or experimentally) well-determined IMFP and an IMFP not accessible either experimentally or observationally may provide a clear path for the  determination of the latter. As expected, the correlations in diagonal blocks $\approx 1$ at all three different densities. At $n=0.15$~fm$^{-3}$, we observe that $\rho(\mathrm{IMFP}_{CC}^{ax},~\mathrm{IMFP}_{NC}^{ax})\approx0.8$. And this high correlation at $n=0.15$~fm$^{-3}$ may result from a relatively high correlation between $g_{nn}$ and $V_\mathrm{gt}$ at this density (see the upper left panel of Fig. \ref{fig:pearsonLFLF}). At $n=0.005$~fm$^{-3}$, we observe relatively high $\rho(\mathrm{IMFP}_{NC}^{ax},~\mathrm{IMFP}_{NC}^{vec})$, $\rho(\mathrm{IMFP}_{NC}^{ax},~\mathrm{IMFP}_{CC}^{vec})$ and $\rho(\mathrm{IMFP}_{NC}^{vec},~\mathrm{IMFP}_{CC}^{vec})$. Again, these strong correlations at off-diagonal blocks may be due to the correlations between EoS-based quantities and EoS-based quantities at $n=0.005$~fm$^{-3}$.

\section{Conclusion}\label{sec:conclusion}
Strong interactions between nucleons could significantly modify neutrino-nucleon cross sections via many-body effects at densities and temperature of relevance to CCSNe. The main difficulty in improving the description of neutrino opacities may come from the poorly-constrained density-dependent nucleon-nucleon interactions. The nucleon-nucleon interactions, also play an important role in determining the properties of EoSs. In this way, the uncertainty of EoS-based quantities propagate to the uncertainty of neutrino opacities.

In this work we calculate the neutrino-nucleon interactions in both vector and axial vector coupling channels, in the framework of the HF+RPA approximation. In the low density regime, the neutrino opacity based on HF+RPA calculations are consistent with a virial EoS, which generates a model-independent nucleon-nucleon interaction (in both spin-independent and spin-dependent channels). The low-density virial EoS used in this work naturally evolve to a series of Skyrme EoSs as density increases, as in Ref.~\cite{Du22hd}. Note that in our framework the neutrino opacity calculations are always consistent with the underlying EoSs at any densities. Although spin-dependent interactions play an important role in determining the dominant axial-vector neutrino-nucleon interactions, they only slightly influence the properties of 
nuclei, which are mostly spin-saturated or close to, and
which are 
employed to fit Skyrme forces.
Consequently, we observe that the EoS-based quantities in spin-dependent channels are not well constrained and have big uncertainties at high densities, which further induce big uncertainties of neutrino opacities in axial-vector channel in our self-consistent calculations. 

In the last several decades, our understanding of nuclear EoSs increases thanks to the progress made in experimental measurements of nuclei properties and in astronomy observations of neutron star properties. These measurements and observations, provide valuable constraints on spin-independent nucleon-nucleon interactions. However, compared to the former, spin-dependent nucleon-nucleon interactions are still poorly constrained. While they play a crucial role in neutrino-nucleon interactions at high densities, in electron capture reactions and in pion condensation. In the future, we will construct: (1) EoSs with extended terms stabilizing spin-dependent functional at high densities \cite{PhysRevC.66.014303}; (2) EoSs that are constrained by recent measurements of Gamow-Teller giant resonance in various finite nuclei \cite{PhysRevLett.121.132501}; and (3) improved connection between the low-density and the high density models, where not only the energy but also the first and second derivatives will be accurately described.

The descriptions of correlations between (1) EoS-based quantities and EoS-based quantities; (2) EoS-based quantities and IMFPs and (3) IMFPs and IMFPs may not be affected by the large uncertainties of EoS-based quantities and IMFPs at high densities. In this work, for the first time we study these correlations in the framework of RPA at different densities. How big are the influence of uncertainties of the EoS-based quantities on neutrino-nucleon IMFPs at different densities? The study of density-dependent correlations may help to answer this question and motivate an accurate determination of EoS-based quantities to better determine neutrino opacities at interested densities in the future.

In the future, it could be interesting to study the impact of neutrino opacity uncertainties on the CCSNe explosion mechanism, on the CCSNe neutrino signals and on the CCSNe light curves.

\subsubsection*{Acknowledgments}
The authors would like to thank J. Navarro and S. Reddy for very helpful discussions. ZL acknowledges funding from the NSF Grant No. PHY-1554876, PHY 21-16686 and from DOE Scidac Grant DE-SC0018232. AWS was supported by NSF AST-1909490 and PHY 21-16686.
JM is supported by CNRS grant PICS-08294 VIPER (Nuclear Physics for Violent Phenomena in the Universe), the CNRS IEA-303083 BEOS project, the CNRS/IN2P3 NewMAC project, and benefit from the LABEX Lyon Institute of Origins (ANR-10-LABX-0066) of the \textsl{Universit\'e Claude Bernard Lyon-1}.

\bibliographystyle{apsrev}
\bibliography{b}

\clearpage
\newpage
\appendix
\section{Density-, Temperature- and Isospin-Dependent Landau Parameters}\label{appendix1}

We express the Skyrme Hamiltonian 
in the following form
\begin{equation}
H_{\mathrm{Sk}} = \frac{k_{Fn}^5} {10 \pi^2 m_n^{*}} + \frac{k_{Fp}^5} {10 \pi^2 m_p^{*}} + H_{\mathrm{pot}}(n_n,n_p),
\end{equation}
where the two first terms are the kinetic terms for the neutrons and the protons with effective mass contribution and $H_{\mathrm{pot}}$ is the potential term  expressed as:
\begin{equation}
    \begin{split}
        H_{\mathrm{pot}}=&\frac{1}{2}n^2t_0\left(1+\frac{x_0}{2}\right)-
        \frac{1}{2}(n_n^2+n_p^2)
        t_0\left(\frac{1}{2}+x_0\right) \\
        &+\frac{1}{12}n^{2+\gamma}t_3 \left(1+\frac{x_3}{2}\right) \\
        & -
        \frac{1}{12}n^\gamma(n_n^2+n_p^2)
        t_3\left(\frac{1}{2}+x_3\right).
    \end{split}
\end{equation}
Let us rewrite the kinetic energy part of $H_{\mathrm{Sk}}$ in terms of the kinetic energy density $\tau_i$ ($i=n$, $p$) defined as:
\begin{equation}
    \tau_i=\frac{3}{5}n_i k_{Fi}^2=\frac{2k_{Fi}^5}{10\pi^2}.
\end{equation}
So, we have:
\begin{equation}
\label{eq:skdensityhamiltonian}
\begin{split}
  H_{\mathrm{Sk}} &= \frac{\tau_n} {2 m_n^{*}} + \frac{\tau_p} 
  {2 m_p^{*}} + H_{\mathrm{pot}}(n_n,n_p)\\&\equiv H_k^n+H_k^p+H_{\mathrm{pot}}(n_n,n_p),
\end{split}
\end{equation}
where we have explicitly expressed the Skyrme Hamiltonian in terms of the independent densities $n_n$, $n_p$, $\tau_n$ and $\tau_p$ in isospin asymmetric matter.
Note that the neutron and proton effective masses are functions of densities as,
\begin{equation}
\begin{split}
    \frac{m_n}{m^*_n}=&1+2m_n\Big\{\frac 1 4 n\big[t_1(1+\frac 1 2 x_1)+t_2(1+\frac 1 2 x_2)\big]+\\& \frac 1 4 n_n\big[-t_1(\frac 1 2+x_1)+t_2(\frac 1 2+x_2)\big]\Big\} \, ,
\end{split}
\end{equation}
and
\begin{equation}
\begin{split}
    \frac{m_p}{m^*_p}=&1+2m_p\Big\{\frac{1}{4} n\Big[t_1(1+1/2x_1)+t_2(1+1/2x_2)\Big]+\\&\frac{1}{4} n_p\Big[-t_1(1/2+x_1)+t_2(1/2+x_2)\Big]\Big\}.
\end{split}
\end{equation}
Given the energy density $H_{\mathrm{Sk}}$ from the Skyrme model, the Landau parameters $f_{nn}$ and $f_{pp}$ are obtained as:
\begin{equation}
\begin{split}
  f_{nn}&=\frac{\partial^2 H_{\mathrm{pot}}}{\partial^2 n_n}+2\frac{\partial^2 H_k^n}{\partial\tau_n\partial n_n}k_{Fn}^2\\&\equiv \frac{1}{2}(W_{1}^{(nn,0)}+W_{1R}^{(nn,0)})+W_2^{(nn,0)}k_{Fn}^2\, ,
\end{split}
\end{equation}
and
\begin{equation}
\begin{split}
    f_{pp}&=\frac{\partial^2 H_{\mathrm{pot}}}{\partial^2 n_p}+2\frac{\partial^2 H_k^p}{\partial\tau_p\partial n_p}k_{Fp}^2\\&\equiv \frac{1}{2}(W_{1}^{(pp,0)}+W_{1R}^{(pp,0)})+W_2^{(pp,0)}k_{Fp}^2.
\end{split}
\end{equation}
Finally, we have: 
\begin{equation}
\begin{split}
    f_{np}&=\frac{\partial^2 
    H_{\mathrm{pot}}}{\partial n_n\partial n_p}+\left(\frac{\partial^2 H_k^p}{\partial\tau_p\partial n_n}k_{Fp}^2+\frac{\partial^2 H_k^n}{\partial\tau_n\partial n_p}k_{Fn}^2\right)\\&\equiv \frac{1}{2}(W_{1}^{(np,0)}+W_{1R}^{(np,0)})+\frac{1}{2}W_2^{(np,0)}(k_{Fn}^2+k_{Fp}^2).
\end{split}
\end{equation}
The strength function $W_i^{(\tau,\tau',0)}$ agrees with Ref.~\cite{Hernanhez:1997zbr}. 

We now discuss the spin-density-dependent Landau parameters in Skyrme model. To begin with, we define the spin density of neutrons and protons as:
\begin{equation}
    n_{n,3}=n_{n\uparrow}-n_{n\downarrow},
\end{equation}
and 
\begin{equation}
    n_{p,3}=n_{p\uparrow}-n_{p\downarrow}.
\end{equation}
We further define the spin-kinetic energy density as:
\begin{equation}
    \tau_{n,3}=\frac{3}{5}n_{n,3}k_{Fn}^{2},
\end{equation}
and \begin{equation}
    \tau_{p,3}=\frac{3}{5}n_{p,3}k_{Fp}^{2}.
\end{equation}
The contribution to the Skyrme Hamiltonian due to the spin asymmetry
is expressed as:
\begin{equation}
    H^s_{\mathrm{Sk}}=H_k^{n,3}+H_k^{p,3}+H_{\mathrm{pot}}^s(n_{n,3},n_{p,3}),
\end{equation}
where the $H_k^{n,3}$, $H_k^{p,3}$ and $H_{\mathrm{pot}}^s(n_{n,3},n_{p,3})$ are defined as following:
\begin{equation}
\begin{split}
    H_k^{n,3}=&\frac 1 8 \tau_{n,3}\Big\{n_{n,3}t_1(-1+x_1)+n_{p,3}t_1x_1+n_{n,3}t_2(1+x_2)\\&+n_{p,3}t_2x_2\Big\}, 
\end{split}
\end{equation}
\begin{equation}
\begin{split}
    H_k^{p,3}=&\frac 1 8 \tau_{p,3}\Big\{n_{p,3}t_1(-1+x_1)+n_{n,3}t_1x_1+n_{p,3}t_2(1+x_2)\\&+n_{n,3}t_2x_2\Big\}, 
\end{split}
\end{equation}
and \begin{equation}
    \begin{split}
        H_{\mathrm{pot}}^s=&-\frac{1}{8}(n_{n,3}-n_{p,3})^2t_0+\frac{1}{8}(n_{n,3}+n_{p,3})^2t_0(-1+2x_0)\\&-\frac{1}{48}n^\gamma(n_{n,3}-n_{p,3})^2t_3\\
        &+\frac{1}{48}n^\gamma(n_{n,3}+n_{p,3})^2t_3(-1+2x_3).
    \end{split}
\end{equation}
Note that in spin-saturated
matter the term $H^s_{\mathrm{Sk}}$ vanishes by construction
and the total Skyrme Hamiltonian reduces to Eq.~\eqref{eq:skdensityhamiltonian}. 
The Landau parameters describing the spin-density fluctuations are expressed as:
\begin{equation}
\begin{split}
    g_{nn}&=\frac{\partial^2 H_{\mathrm{pot}}^s}{\partial^2 n_{n,3}}+2\frac{\partial^2 H_k^{n,3}}{\partial\tau_{n,3}\partial n_{n,3}}k_{Fn}^2\\&\equiv \frac{1}{2}W_{1}^{(nn,1)}+W_2^{(nn,1)}k_{Fn}^2,
\end{split}
\end{equation}
\begin{equation}
\begin{split}
    g_{pp}&=\frac{\partial^2 H_{\mathrm{pot}}^s}{\partial^2 n_{p,3}}+2\frac{\partial^2 H_k^{p,3}}{\partial\tau_{p,3}\partial n_{p,3}}k_{Fp}^2\\&\equiv \frac{1}{2}W_{1}^{(pp,1)}+W_2^{(pp,1)}k_{Fn}^2,
\end{split}
\end{equation}
and \begin{equation}
\begin{split}
    g_{np}&=\frac{\partial^2 H_{\mathrm{pot}}^s}{\partial n_{n,3}\partial n_{p,3}}+\left(\frac{\partial^2 H_k^{n,3}}{\partial\tau_{n,3}\partial n_{p,3}}k_{Fn}^2+\frac{\partial^2 H_k^{p,3}}{\partial\tau_{p,3}\partial n_{n,3}}k_{Fp}^2\right)\\&\equiv \frac{1}{2}W_{1}^{(np,1)}+\frac{1}{2}W_2^{(np,1)}(k_{Fn}^2+k_{Fp}^2).
\end{split}
\end{equation}
The strength function of the residual interactions $W_i^{(\tau\tau',1)}$ agrees with Ref.~\cite{Hernanhez:1997zbr}. 

We then discuss the vector and the axial vector Landau parameters in CC reactions $n+\bar{\nu}_e\rightarrow p+e^-$ based on Skyrme models. For $V_\mathrm{f}$, we have:
\begin{equation}
    V_\mathrm{f}=\frac{1}{2}(W_1^{(nn,0)}-W_1^{(np,0)})+(W_2^{(nn,0)}-W_2^{(np,0)})k^2_{Fn},
\end{equation}
where $W_1^{(\tau\tau',0)}$ are the strength of Landau parameters.
Note the absence of the rearrangement terms in the CC channel since the density dependent term of the Skyrme interaction does not contribute here.
Here the rearrangement term $W_{1R}^{(\tau\tau',0)}$ is:
\begin{equation}
    W_{1R}^{(nn,0)}=\frac{2t_3}{3}\gamma n^{\gamma-1}\left[n(1+x_3/2)-n_n(x_3+1/2)\right],
\end{equation}
and
\begin{equation}
    W_{1R}^{(np,0)}=\frac{1}{2}\gamma t_3n^{\gamma}.
\end{equation}
There are no rearrangement contributions to spin-density dependent residual interactions and we have:
\begin{equation}
    V_\mathrm{gt}=\frac{1}{2}(W_1^{(nn,1)}-W_1^{(np,1)})+(W_2^{(nn,1)}-W_2^{(np,1)})k^2_{Fn}.
\end{equation}
In this work Landau parameters in the high density regime are derived based on Skyrme models, while in the low density regime they are derived based on model-independent virial interactions. In the low density region where the virial interaction applies, we assume that $m^*\approx m$, and the kinetic energy density terms in the virial Hamiltonian is density-independent. In the first-order approximation where only the 1st order virial coefficients are involved, the potential energy density part (including both density and spin-density terms) of the virial Hamiltonian is:
\begin{equation}
\begin{split}
    H_{\mathrm{pot}}^{\mathrm{virial}}=
    &-\frac{1}{2}\hat{b}_{pn,0}T\lambda^3
    \left[(n_n+n_{n,3})(n_p-n_{p,3})\right.\\
    &\left.+(n_n-n_{n,3})(n_p+n_{p,3})\right]\\
    &-\frac{1}{2}\hat{b}_{pn,1}T\lambda^3
    \left[(n_n+n_{n,3})(n_p+n_{p,3})\right.\\
    &\left.+(n_n+n_{n,3})(n_p+n_{p,3})\right]\\
    &-\frac{1}{2}\hat{b}_{n,0}T\lambda^3
    \left[(n_n+n_{n,3})(n_n-n_{n,3})\right.\\
    &\left.+(n_p+n_{p,3})(n_p-n_{p,3})\right]\\
    &-\frac{1}{4}\hat{b}_{n,1}T\lambda^3
    \left[(n_n+n_{n,3})^2+(n_n-n_{n,3})^2\right.\\
    &\left.+(n_p+n_{p,3})^2+(n_p-n_{p,3})^2\right] \, ,
\end{split}
\end{equation}
where the notation of virial coefficients follows Refs.~\cite{Horowitz:2016gul,Horowitz:2012us}.
Since in the virial Hamiltonian the kinetic energy density is density independent. So, we have $H_k^{n,\mathrm{virial}}=\tau_n/(2m_n)$ and $H_k^{p,virial}=\tau_p/(2m_p)$. The Landau parameters in virial model simplifies to:
\begin{equation}
    f_{nn}^{\mathrm{virial}}=\frac{\partial^2 H_{\mathrm{pot}}^{\mathrm{virial}}}{\partial^2 n_n},
\end{equation}
\begin{equation}
    f_{pp}^{\mathrm{virial}}=\frac{\partial^2 H_{\mathrm{pot}}^{\mathrm{virial}}}{\partial^2 n_p},
\end{equation}
\begin{equation}
    f_{np}^{\mathrm{virial}}=\frac{\partial^2 H_{\mathrm{pot}}^{\mathrm{virial}}}{\partial n_n\partial n_p},
\end{equation}
\begin{equation}
    g_{nn}^{\mathrm{virial}}=\frac{\partial^2 H_{\mathrm{pot}}^{\mathrm{virial}}}{\partial^2 n_{n,3}},
\end{equation}
\begin{equation}
    g_{pp}^{\mathrm{virial}}=\frac{\partial^2 H_{\mathrm{pot}}^{\mathrm{virial}}}{\partial^2 n_{p,3}},
\end{equation}
\begin{equation}
    g_{np}^{\mathrm{virial}}=\frac{\partial^2 H_{\mathrm{pot}}^{\mathrm{virial}}}{\partial n_{n,3}\partial n_{p,3}}.
\end{equation}
Note that in $H_{\mathrm{pot}}^{\mathrm{virial}}$ the density dependent terms include up to $\mathcal{O}(n^2)$, which means that the virial Landau parameters are density independent in the first order of approximation. So, there are no rearrangement contributions in virial Landau parameters and the Landau parameters in CC interaction based on virial method is: 
\begin{equation}
    V_\mathrm{f,virial}=f_{nn}^{\mathrm{virial}}-f_{np}^{\mathrm{virial}},
\end{equation}
and \begin{equation}
    V_\mathrm{gt,virial}=g_{nn}^{\mathrm{virial}}-g_{np}^{\mathrm{virial}}.
\end{equation}
Finally, we discuss the Landau parameters that apply in the whole density regime. As described above, we construct
a global Hamiltonian $H_{g}$ that applies in the whole density regime:
\begin{equation}
    H_{\eta}=[1-\eta(z_n,z_p)]H_{Sk}+\eta(z_n,z_p)H_{\mathrm{virial}}
\end{equation}
where $\eta(z_n,z_p)$ is given in Eq.~\ref{eq:gfunc}.
For the general density-dependent Landau parameters in NC reaction, we then have:
\begin{equation}\label{eq:fnngeneral}
\begin{split}
     f_{nn}^{\eta}=&\eta f_{nn}^{\mathrm{virial}}+\frac{1}{2}(1-\eta)(W_{1}^{(nn,0)}+W_{1R}^{(nn,0)})+
     W_{1,\eta}^{(nn,0)}\\&+(1-\eta)W_2^{(nn,0)}\times k_{Fn}^{2}+W_{2,\eta}^{(nn,0)}\times k_{Fn}^2\\&\equiv \frac{1}{2}W_{1,\mathrm{general}}^{(nn,0)}+W_{2,\mathrm{general}}^{(nn,0)}k_{Fn}^2,
\end{split}
\end{equation}
where \begin{equation}
    W_{1,\eta}^{(nn,0)}=\frac{\partial^2 \eta}{\partial^2n_n}(H_{\mathrm{pot}}^{\mathrm{virial}}-H_{\mathrm{pot}})+2\frac{\partial \eta}{\partial n_n}\frac{\partial(H_{\mathrm{pot}}^{\mathrm{virial}}-H_{\mathrm{pot}})}{\partial n_n}
\end{equation}
and \begin{equation}
    W_{2,\eta}^{(nn,0)}=2\frac{\partial \eta}{\partial n_n}\frac{\partial(H_k^{n,\mathrm{virial}}-H_k^n)}{\partial \tau_n}
\end{equation}
are the corrections to Landau parameter strength in vector channel due to the fact that \emph{$\eta$ is density dependent}. Note that the corrections are proportional to $H^{\mathrm{virial}}-H$. So, the corrections are small when the virial and the Skyrme Hamiltonian approximately agree.
$f_{pp}^{\eta}$ is derived similarly as $f_{nn}^{\eta}$, with $n\leftrightarrow p$. We also have: 
\begin{equation}\label{eq:fnpgeneral}
\begin{split}
     f_{np}^{\eta}=&\eta f_{np}^{\mathrm{virial}}+\frac{1}{2}(1-\eta)(W_{1}^{(np,0)}+W_{1R}^{(np,0)})+W_{1,\eta}^{(np,0)}\\&+\frac{1}{2}(1-\eta)W_2^{(np,0)}(k_{Fn}^2+k_{Fp}^2)\\&+\frac{1}{2}W_{2,\eta n}^{(np,0)}k_{Fn}^2+\frac{1}{2}W_{2,\eta p}^{(np,0)}k_{Fp}^2\\&\equiv \frac{1}{2}W_{1,\mathrm{general}}^{(np,0)}+\frac{1}{2}W_{2,\mathrm{general}}^{(np,0)}(k_{Fn}^2+k_{Fp}^2),
\end{split}
\end{equation}
where \begin{equation}
\begin{split}
    W_{1,\eta}^{(np,0)}&=\frac{\partial^2 \eta}{\partial n_n\partial n_p}(H_{\mathrm{pot}}^{\mathrm{virial}}-H_{\mathrm{pot}})\\&+\frac{\partial \eta}{\partial n_n}\frac{\partial(H_{\mathrm{pot}}^{\mathrm{virial}}-H_{\mathrm{pot}})}{\partial n_p}+\frac{\partial \eta}{\partial n_p}\frac{\partial(H_{\mathrm{pot}}^{\mathrm{virial}}-H_{\mathrm{pot}})}{\partial n_n},
\end{split}
\end{equation}
\begin{equation}
    W_{2,\eta n}^{(np,0)}=2\frac{\partial \eta}{\partial n_p}\frac{\partial(H_k^{n,\mathrm{virial}}-H_k^n)}{\partial \tau_n}
\end{equation}
and \begin{equation}
    W_{2,\eta p}^{(np,0)}=2\frac{\partial \eta}{\partial n_n}\frac{\partial(H_k^{p,\mathrm{virial}}-H_k^p)}{\partial \tau_p}
\end{equation}
are the corrections to Laudau parameter strength in axial channel due to the fact that \emph{$\eta$ is density dependent}.
Since the transition function g is spin-density independent, the global spin-density dependent Landau parameter simplifies to:
\begin{equation}\label{eq:gnngeneral}
\begin{split}
    g_{nn}^{\eta}&=\eta g_{nn}^{\mathrm{virial}}+\frac{1}{2}(1-\eta)W_1^{(nn,1)}+(1-\eta)W_2^{(nn,1)}\times k_{Fn}^{2}\\&\equiv \frac{1}{2}W_{1,\mathrm{general}}^{(nn,1)}+W_{2,\mathrm{general}}^{(nn,1)}k_{Fn}^2
\end{split}
\end{equation}
\begin{equation}\label{eq:gnpgeneral}
\begin{split}
    g_{np}^{\eta}&=\eta g_{np}^{\mathrm{virial}}+\frac{1}{2}(1-\eta)W_1^{(np,1)}+\frac{1}{2}(1-\eta)W_2^{(np,1)}\\&\times (k_{Fn}^{2}+k_{Fp}^2)\\&\equiv \frac{1}{2}W_{1,\mathrm{general}}^{(np,1)}+\frac{1}{2}W_{2,\mathrm{general}}^{(np,1)}(k_{Fn}^2+k_{Fp}^2).
\end{split}
\end{equation}
For CC interaction, again we need to take off terms due to rearrangement effects in Hamiltonian and single particle potentials. We derive the rearrangement term in single particle potential $U\equiv \partial H_{\mathrm{pot}}/\partial n$ in Skyrme model being:
\begin{equation}
    U^{\mathrm{Re}}=\frac{\alpha t_3}{12}n^{\alpha-1}((1+x_3/2)n^2-(x_3+1/2)(n_n^2+n_p^2)).
\end{equation}
We then have:
\begin{equation}\label{eq:vfgeneral}
\begin{split}
    V_{\mathrm{f}}^\eta&=\eta(f_{nn}^{\mathrm{virial}}-f_{np}^{\mathrm{virial}})+\frac{1}{2}(1-\eta)(W_1^{(nn,0)}-W_1^{(np,0)})\\&+[W_{1,\eta}^{(nn,0)}-W_{1,\eta}^{(np,0)}+2\frac{\partial \eta}{\partial n_n}U^{\mathrm{Re}}-(\frac{\partial \eta}{\partial n_n}+\frac{\partial \eta}{\partial n_p})U^{\mathrm{Re}}]\\&+[(1-\eta)(W_2^{(nn,0)}-W_2^{(np,0)})+W_{2,\eta}^{(nn,0)}\\&-W_{2,\eta n}^{(np,0)}\frac{k_{Fn}^2}{k_{Fn}^2+k_{Fp}^2}-W_{2,\eta p}^{(np,0)}\frac{k_{Fp}^2}{k_{Fn}^2+k_{Fp}^2}]k_{Fn}^2\\&\equiv \frac{1}{2}(\tilde{W}_{1,\mathrm{general}}^{(nn,0)}-\tilde{W}_{1,\mathrm{general}}^{(np,0)})+[2\frac{\partial \eta}{\partial n_n}U^{\mathrm{Re}}\\&-(\frac{\partial \eta}{\partial n_n}+\frac{\partial \eta}{\partial n_p})U^{\mathrm{Re}}]+(W_{2,\eta}^{(nn,0)}-W_{2,\eta}^{(np,0)})k_{Fn}^2,
\end{split}
\end{equation}
where $\tilde{W}_{1,\mathrm{general}}^{(\tau\tau',0)}$ is obtained by replacing $W_{1}^{(\tau\tau',0)}+W_{1R}^{(\tau\tau',0)}\rightarrow W_1^{(\tau\tau',0)}$ in $W_{1,\mathrm{general}}^{(\tau\tau',0)}$. And the second bracket with $U^{\mathrm{Re}}$ is due to the removal of rearrangement contributions in $W_{1,\eta}^{(\tau,\tau',0)}$. 
and 
\begin{equation}\label{eq:vgtgeneral}
    \begin{split}
        V_{\mathrm{gt}}^\eta=&\eta(g_{nn}^{\mathrm{virial}}-g_{np}^{\mathrm{virial}})+\frac{1}{2}(1-\eta)(W_1^{(nn,1)}-W_1^{(np,1)})\\&+(1-\eta)(W_2^{(nn,1)}-W_2^{(np,1)})k_{Fn}^2\\&\equiv \frac{1}{2}(W_{1,\mathrm{general}}^{(nn,1)}-W_{1,\mathrm{general}}^{(np,1)})\\&+(W_{2,\mathrm{general}}^{(nn,1)}-W_{2,\mathrm{general}}^{(np,1)})k_{Fn}^2.
    \end{split}
\end{equation}

The $\eta$ function was first applied in \cite{PhysRevC.99.025803} to construct a Hamiltonian valid from low-density regime to high-density regime. This $\eta$ function ensures that in low-density regime the EoS in \cite{PhysRevC.99.025803} reproduces the features predicted by a virial EoS and allows the EoS to smoothly transit into a Skyrme EoS as the density increases. However, the location of a transition region where the $\eta$ function smoothly decrease from $\approx1$ to $\approx0$ may be model-dependent, and in principle, temperature-dependent as well. A precise connection between the low and the high density regimes shall be performed in the future, ensuring that this connection does not impact the first and second derivatives of the energy density. Since this is not performed in the present study, the functional form
of the $\eta$ function in this transition region may influence the behavior of $W_{1,\eta}$, $W_{2,\eta n}$, $W_{2,\eta p}$, $\frac{\partial \eta}{\partial n_n}U^{\mathrm{Re}}$ and $\frac{\partial \eta}{\partial n_p}U^{\mathrm{Re}}$.
In the present study, we simply neglect the contribution of the derivatives of the $\eta$ function to the Landau parameter strengths for simplicity. This shall however be investigated in more details in the future. 

\section{Landau Parameter Strength}\label{appendix2}
Here we summarized the strength of Landau parameters $W_i^{\tau\tau',S}$. The form of these Laudau parameter strengths agree with that in Ref.~\cite{Hernanhez:1997zbr}. Note that our $W_{1}^{\tau\tau/\tau\tau',0}+W_{1R}^{\tau\tau/\tau\tau',0}$ is equivalent to the $W_1^{\tau\tau/\tau\tau',0}$ terms in Ref.~\cite{Hernanhez:1997zbr}, which include the contribution from rearrangement terms.
\begin{equation}
\begin{split}
     W_{1}^{\tau\tau,0}+W_{1R}^{\tau\tau,0}&=t_0(1-x_0)+\frac{1}{6}t_3\rho^\gamma(1-x_3)\\&+\frac{2}{3}\gamma t_3\rho^{\gamma-1}[(1+\frac{1}{2}x_3)\rho-(\frac{1}{2}+x_3)\rho_\tau]\\&+\frac{1}{6}\gamma(\gamma-1)t_3\rho^{\gamma-2}[(1+\frac{1}{2}x_3)\rho^2-(\frac{1}{2}+x_3)(\rho_n^2+\rho_p^2)],
\end{split}
\end{equation}

\begin{equation}
\begin{split}
     W_2^{\tau\tau,0}&=\frac{1}{4}[t_1(1-x_1)+3t_2(1+x_2)],
\end{split}
\end{equation}

\begin{equation}
\begin{split}
     W_{1}^{\tau\tau,1}&=t_0(x_0-1)+\frac{1}{6}\rho^\gamma t_3(x_3-1),
\end{split}
\end{equation}

\begin{equation}
\begin{split}
     W_2^{\tau\tau,1}&=\frac{1}{4}[t_1(x_1-1)+t_2(1+x_2)],
\end{split}
\end{equation}

\begin{equation}
\begin{split}
     W_{1}^{\tau-\tau,0}+W_{1R}^{\tau-\tau,0}&=t_0(2+x_0)+\frac{1}{6}t_3\rho^\gamma(2+x_3)+\frac{1}{2}\gamma t_3\rho^\gamma\\&+\frac{1}{6}\gamma(\gamma-1)t_3\rho^{\gamma-2}[(1+\frac{1}{2}x_3)\rho^2\\&-(\frac{1}{2}+x_3)(\rho^2_n+\rho_p^2)],
\end{split}
\end{equation}

\begin{equation}
\begin{split}
     W_2^{\tau-\tau,0}&=\frac{1}{4}[t_1(2+x_1)+t_2(2+x_2)],
\end{split}
\end{equation}

\begin{equation}
\begin{split}
     W_{1}^{\tau-\tau,1}&=t_0x_0+\frac{1}{6}t_3\rho^\gamma x_3,
\end{split}
\end{equation}

\begin{equation}
\begin{split}
     W_2^{\tau-\tau,1}&=\frac{1}{4}(t_1x_1+t_2x_2).
\end{split}
\end{equation}


\end{document}